\documentclass[aps,prl,twocolumn]{revtex4-1}  
\usepackage{graphicx}  
\usepackage{dcolumn}   
\usepackage{bm}        
\usepackage{amssymb}   
\usepackage{color} %
\RequirePackage{xspace}
\usepackage{relsize}
\usepackage{colortbl}
\usepackage{amsmath} 
\usepackage{ifthen} 
\usepackage{subfigure}

\usepackage{ifthen} 
\graphicspath{{ps}}
\newboolean{pdflatex}
\setboolean{pdflatex}{true} 
\usepackage{rotating} 

\newboolean{articletitles}
\setboolean{articletitles}{true} 

\newboolean{uprightparticles}
\setboolean{uprightparticles}{false} 

\usepackage{amssymb}
\usepackage{amsfonts}
\usepackage{upgreek} 

\usepackage{hyperref}    
\usepackage[all]{hypcap} 
\begin{document}



\def\lhcb {LHCb\xspace}
\def\ux85 {UX85\xspace}
\def\cern {CERN\xspace}
\def\lhc {LHC\xspace}
\def\atlas {ATLAS\xspace}
\def\cms {CMS\xspace}
\def\babar  {BaBar\xspace}
\def\belle  {Belle\xspace}
\def\aleph  {ALEPH\xspace}
\def\delphi {DELPHI\xspace}
\def\opal   {OPAL\xspace}
\def\lthree {L3\xspace}
\def\lep    {LEP\xspace}
\def\cdf    {CDF\xspace}
\def\dzero  {D\O\xspace}
\def\sld    {SLD\xspace}
\def\cleo   {CLEO\xspace}
\def\uaone  {UA1\xspace}
\def\uatwo  {UA2\xspace}
\def\tevatron {TEVATRON\xspace}


\def\pu     {PU\xspace}
\def\velo   {VELO\xspace}
\def\rich   {RICH\xspace}
\def\richone {RICH1\xspace}
\def\richtwo {RICH2\xspace}
\def\ttracker {TT\xspace}
\def\intr   {IT\xspace}
\def\st     {ST\xspace}
\def\ot     {OT\xspace}
\def\Tone   {T1\xspace}
\def\Ttwo   {T2\xspace}
\def\Tthree {T3\xspace}
\def\Mone   {M1\xspace}
\def\Mtwo   {M2\xspace}
\def\Mthree {M3\xspace}
\def\Mfour  {M4\xspace}
\def\Mfive  {M5\xspace}
\def\ecal   {ECAL\xspace}
\def\spd    {SPD\xspace}
\def\presh  {PS\xspace}
\def\hcal   {HCAL\xspace}
\def\bcm    {BCM\xspace}

\def\ode    {ODE\xspace}
\def\daq    {DAQ\xspace}
\def\tfc    {TFC\xspace}
\def\ecs    {ECS\xspace}
\def\lone   {L0\xspace}
\def\hlt    {HLT\xspace}
\def\hltone {HLT1\xspace}
\def\hlttwo {HLT2\xspace}


\ifthenelse{\boolean{uprightparticles}}%
{\def\Palpha      {\ensuremath{\upalpha}\xspace}
 \def\Pbeta       {\ensuremath{\upbeta}\xspace}
 \def\Pgamma      {\ensuremath{\upgamma}\xspace}                 
 \def\Pdelta      {\ensuremath{\updelta}\xspace}                 
 \def\Pepsilon    {\ensuremath{\upepsilon}\xspace}                 
 \def\Pvarepsilon {\ensuremath{\upvarepsilon}\xspace}                 
 \def\Pzeta       {\ensuremath{\upzeta}\xspace}                 
 \def\Peta        {\ensuremath{\upeta}\xspace}                 
 \def\Ptheta      {\ensuremath{\uptheta}\xspace}                 
 \def\Pvartheta   {\ensuremath{\upvartheta}\xspace}                 
 \def\Piota       {\ensuremath{\upiota}\xspace}                 
 \def\Pkappa      {\ensuremath{\upkappa}\xspace}                 
 \def\Plambda     {\ensuremath{\uplambda}\xspace}                 
 \def\Pmu         {\ensuremath{\upmu}\xspace}                 
 \def\Pnu         {\ensuremath{\upnu}\xspace}                 
 \def\Pxi         {\ensuremath{\upxi}\xspace}                 
 \def\Ppi         {\ensuremath{\uppi}\xspace}                 
 \def\Pvarpi      {\ensuremath{\upvarpi}\xspace}                 
 \def\Prho        {\ensuremath{\uprho}\xspace}                 
 \def\Pvarrho     {\ensuremath{\upvarrho}\xspace}                 
 \def\Ptau        {\ensuremath{\uptau}\xspace}                 
 \def\Pupsilon    {\ensuremath{\upupsilon}\xspace}                 
 \def\Pphi        {\ensuremath{\upphi}\xspace}                 
 \def\Pvarphi     {\ensuremath{\upvarphi}\xspace}                 
 \def\Pchi        {\ensuremath{\upchi}\xspace}                 
 \def\Ppsi        {\ensuremath{\uppsi}\xspace}                 
 \def\Pomega      {\ensuremath{\upomega}\xspace}                 

 \def\PDelta      {\ensuremath{\Delta}\xspace}                 
 \def\PXi      {\ensuremath{\Xi}\xspace}                 
 \def\PLambda      {\ensuremath{\Lambda}\xspace}                 
 \def\PSigma      {\ensuremath{\Sigma}\xspace}                 
 \def\POmega      {\ensuremath{\Omega}\xspace}                 
 \def\PUpsilon      {\ensuremath{\Upsilon}\xspace}                 
 

 \def\PA      {\ensuremath{\mathrm{A}}\xspace}                 
 \def\PB      {\ensuremath{\mathrm{B}}\xspace}                 
 \def\PC      {\ensuremath{\mathrm{C}}\xspace}                 
 \def\PD      {\ensuremath{\mathrm{D}}\xspace}                 
 \def\PE      {\ensuremath{\mathrm{E}}\xspace}                 
 \def\PF      {\ensuremath{\mathrm{F}}\xspace}                 
 \def\PG      {\ensuremath{\mathrm{G}}\xspace}                 
 \def\PH      {\ensuremath{\mathrm{H}}\xspace}                 
 \def\PI      {\ensuremath{\mathrm{I}}\xspace}                 
 \def\PJ      {\ensuremath{\mathrm{J}}\xspace}                 
 \def\PK      {\ensuremath{\mathrm{K}}\xspace}                 
 \def\PL      {\ensuremath{\mathrm{L}}\xspace}                 
 \def\PM      {\ensuremath{\mathrm{M}}\xspace}                 
 \def\PN      {\ensuremath{\mathrm{N}}\xspace}                 
 \def\PO      {\ensuremath{\mathrm{O}}\xspace}                 
 \def\PP      {\ensuremath{\mathrm{P}}\xspace}                 
 \def\PQ      {\ensuremath{\mathrm{Q}}\xspace}                 
 \def\PR      {\ensuremath{\mathrm{R}}\xspace}                 
 \def\PS      {\ensuremath{\mathrm{S}}\xspace}                 
 \def\PT      {\ensuremath{\mathrm{T}}\xspace}                 
 \def\PU      {\ensuremath{\mathrm{U}}\xspace}                 
 \def\PV      {\ensuremath{\mathrm{V}}\xspace}                 
 \def\PW      {\ensuremath{\mathrm{W}}\xspace}                 
 \def\PX      {\ensuremath{\mathrm{X}}\xspace}                 
 \def\PY      {\ensuremath{\mathrm{Y}}\xspace}                 
 \def\PZ      {\ensuremath{\mathrm{Z}}\xspace}                 
 \def\Pa      {\ensuremath{\mathrm{a}}\xspace}                 
 \def\Pb      {\ensuremath{\mathrm{b}}\xspace}                 
 \def\Pc      {\ensuremath{\mathrm{c}}\xspace}                 
 \def\Pd      {\ensuremath{\mathrm{d}}\xspace}                 
 \def\Pe      {\ensuremath{\mathrm{e}}\xspace}                 
 \def\Pf      {\ensuremath{\mathrm{f}}\xspace}                 
 \def\Pg      {\ensuremath{\mathrm{g}}\xspace}                 
 \def\Ph      {\ensuremath{\mathrm{h}}\xspace}                 
 \def\Pi      {\ensuremath{\mathrm{i}}\xspace}                 
 \def\Pj      {\ensuremath{\mathrm{j}}\xspace}                 
 \def\Pk      {\ensuremath{\mathrm{k}}\xspace}                 
 \def\Pl      {\ensuremath{\mathrm{l}}\xspace}                 
 \def\Pm      {\ensuremath{\mathrm{m}}\xspace}                 
 \def\Pn      {\ensuremath{\mathrm{n}}\xspace}                 
 \def\Po      {\ensuremath{\mathrm{o}}\xspace}                 
 \def\Pp      {\ensuremath{\mathrm{p}}\xspace}                 
 \def\Pq      {\ensuremath{\mathrm{q}}\xspace}                 
 \def\Pr      {\ensuremath{\mathrm{r}}\xspace}                 
 \def\Ps      {\ensuremath{\mathrm{s}}\xspace}                 
 \def\Pt      {\ensuremath{\mathrm{t}}\xspace}                 
 \def\Pu      {\ensuremath{\mathrm{u}}\xspace}                 
 \def\Pv      {\ensuremath{\mathrm{v}}\xspace}                 
 \def\Pw      {\ensuremath{\mathrm{w}}\xspace}                 
 \def\Px      {\ensuremath{\mathrm{x}}\xspace}                 
 \def\Py      {\ensuremath{\mathrm{y}}\xspace}                 
 \def\Pz      {\ensuremath{\mathrm{z}}\xspace}                 
}
{\def\Palpha      {\ensuremath{\alpha}\xspace}
 \def\Pbeta       {\ensuremath{\beta}\xspace}
 \def\Pgamma      {\ensuremath{\gamma}\xspace}                 
 \def\Pdelta      {\ensuremath{\delta}\xspace}                 
 \def\Pepsilon    {\ensuremath{\epsilon}\xspace}                 
 \def\Pvarepsilon {\ensuremath{\varepsilon}\xspace}                 
 \def\Pzeta       {\ensuremath{\zeta}\xspace}                 
 \def\Peta        {\ensuremath{\eta}\xspace}                 
 \def\Ptheta      {\ensuremath{\theta}\xspace}                 
 \def\Pvartheta   {\ensuremath{\vartheta}\xspace}                 
 \def\Piota       {\ensuremath{\iota}\xspace}                 
 \def\Pkappa      {\ensuremath{\kappa}\xspace}                 
 \def\Plambda     {\ensuremath{\lambda}\xspace}                 
 \def\Pmu         {\ensuremath{\mu}\xspace}                 
 \def\Pnu         {\ensuremath{\nu}\xspace}                 
 \def\Pxi         {\ensuremath{\xi}\xspace}                 
 \def\Ppi         {\ensuremath{\pi}\xspace}                 
 \def\Pvarpi      {\ensuremath{\varpi}\xspace}                 
 \def\Prho        {\ensuremath{\rho}\xspace}                 
 \def\Pvarrho     {\ensuremath{\varrho}\xspace}                 
 \def\Ptau        {\ensuremath{\tau}\xspace}                 
 \def\Pupsilon    {\ensuremath{\upsilon}\xspace}                 
 \def\Pphi        {\ensuremath{\phi}\xspace}                 
 \def\Pvarphi     {\ensuremath{\varphi}\xspace}                 
 \def\Pchi        {\ensuremath{\chi}\xspace}                 
 \def\Ppsi        {\ensuremath{\psi}\xspace}                 
 \def\Pomega      {\ensuremath{\omega}\xspace}                 
 \mathchardef\PDelta="7101
 \mathchardef\PXi="7104
 \mathchardef\PLambda="7103
 \mathchardef\PSigma="7106
 \mathchardef\POmega="710A
 \mathchardef\PUpsilon="7107
 \def\PA      {\ensuremath{A}\xspace}                 
 \def\PB      {\ensuremath{B}\xspace}                 
 \def\PC      {\ensuremath{C}\xspace}                 
 \def\PD      {\ensuremath{D}\xspace}                 
 \def\PE      {\ensuremath{E}\xspace}                 
 \def\PF      {\ensuremath{F}\xspace}                 
 \def\PG      {\ensuremath{G}\xspace}                 
 \def\PH      {\ensuremath{H}\xspace}                 
 \def\PI      {\ensuremath{I}\xspace}                 
 \def\PJ      {\ensuremath{J}\xspace}                 
 \def\PK      {\ensuremath{K}\xspace}                 
 \def\PL      {\ensuremath{L}\xspace}                 
 \def\PM      {\ensuremath{M}\xspace}                 
 \def\PN      {\ensuremath{N}\xspace}                 
 \def\PO      {\ensuremath{O}\xspace}                 
 \def\PP      {\ensuremath{P}\xspace}                 
 \def\PQ      {\ensuremath{Q}\xspace}                 
 \def\PR      {\ensuremath{R}\xspace}                 
 \def\PS      {\ensuremath{S}\xspace}                 
 \def\PT      {\ensuremath{T}\xspace}                 
 \def\PU      {\ensuremath{U}\xspace}                 
 \def\PV      {\ensuremath{V}\xspace}                 
 \def\PW      {\ensuremath{W}\xspace}                 
 \def\PX      {\ensuremath{X}\xspace}                 
 \def\PY      {\ensuremath{Y}\xspace}                 
 \def\PZ      {\ensuremath{Z}\xspace}                 
 \def\Pa      {\ensuremath{a}\xspace}                 
 \def\Pb      {\ensuremath{b}\xspace}                 
 \def\Pc      {\ensuremath{c}\xspace}                 
 \def\Pd      {\ensuremath{d}\xspace}                 
 \def\Pe      {\ensuremath{e}\xspace}                 
 \def\Pf      {\ensuremath{f}\xspace}                 
 \def\Pg      {\ensuremath{g}\xspace}                 
 \def\Ph      {\ensuremath{h}\xspace}                 
 \def\Pi      {\ensuremath{i}\xspace}                 
 \def\Pj      {\ensuremath{j}\xspace}                 
 \def\Pk      {\ensuremath{k}\xspace}                 
 \def\Pl      {\ensuremath{l}\xspace}                 
 \def\Pm      {\ensuremath{m}\xspace}                 
 \def\Pn      {\ensuremath{n}\xspace}                 
 \def\Po      {\ensuremath{o}\xspace}                 
 \def\Pp      {\ensuremath{p}\xspace}                 
 \def\Pq      {\ensuremath{q}\xspace}                 
 \def\Pr      {\ensuremath{r}\xspace}                 
 \def\Ps      {\ensuremath{s}\xspace}                 
 \def\Pt      {\ensuremath{t}\xspace}                 
 \def\Pu      {\ensuremath{u}\xspace}                 
 \def\Pv      {\ensuremath{v}\xspace}                 
 \def\Pw      {\ensuremath{w}\xspace}                 
 \def\Px      {\ensuremath{x}\xspace}                 
 \def\Py      {\ensuremath{y}\xspace}                 
 \def\Pz      {\ensuremath{z}\xspace}                 
}



\let\emi\en
\def\electron   {\ensuremath{\Pe}\xspace}
\def\en         {\ensuremath{\Pe^-}\xspace}   
\def\ep         {\ensuremath{\Pe^+}\xspace}
\def\epm        {\ensuremath{\Pe^\pm}\xspace} 
\def\epem       {\ensuremath{\Pe^+\Pe^-}\xspace}
\def\ee         {\ensuremath{\Pe^-\Pe^-}\xspace}

\def\mmu        {\ensuremath{\Pmu}\xspace}
\def\mup        {\ensuremath{\Pmu^+}\xspace}
\def\mun        {\ensuremath{\Pmu^-}\xspace} 
\def\mumu       {\ensuremath{\Pmu^+\Pmu^-}\xspace}
\def\mtau       {\ensuremath{\Ptau}\xspace}

\def\taup       {\ensuremath{\Ptau^+}\xspace}
\def\taum       {\ensuremath{\Ptau^-}\xspace}
\def\tautau     {\ensuremath{\Ptau^+\Ptau^-}\xspace}

\def\ellm       {\ensuremath{\ell^-}\xspace}
\def\ellp       {\ensuremath{\ell^+}\xspace}
\def\ellell     {\ensuremath{\ell^+ \ell^-}\xspace}

\def\neu        {\ensuremath{\Pnu}\xspace}
\def\neub       {\ensuremath{\overline{\Pnu}}\xspace}
\def\nuenueb    {\ensuremath{\neu\neub}\xspace}
\def\neue       {\ensuremath{\neu_e}\xspace}
\def\neueb      {\ensuremath{\neub_e}\xspace}
\def\neueneueb  {\ensuremath{\neue\neueb}\xspace}
\def\neum       {\ensuremath{\neu_\mu}\xspace}
\def\neumb      {\ensuremath{\neub_\mu}\xspace}
\def\neumneumb  {\ensuremath{\neum\neumb}\xspace}
\def\neut       {\ensuremath{\neu_\tau}\xspace}
\def\neutb      {\ensuremath{\neub_\tau}\xspace}
\def\neutneutb  {\ensuremath{\neut\neutb}\xspace}
\def\neul       {\ensuremath{\neu_\ell}\xspace}
\def\neulb      {\ensuremath{\neub_\ell}\xspace}
\def\neulneulb  {\ensuremath{\neul\neulb}\xspace}


\def\g      {\ensuremath{\Pgamma}\xspace}
\def\H      {\ensuremath{\PH^0}\xspace}
\def\Hp     {\ensuremath{\PH^+}\xspace}
\def\Hm     {\ensuremath{\PH^-}\xspace}
\def\Hpm    {\ensuremath{\PH^\pm}\xspace}
\def\W      {\ensuremath{\PW}\xspace}
\def\Wp     {\ensuremath{\PW^+}\xspace}
\def\Wm     {\ensuremath{\PW^-}\xspace}
\def\Wpm    {\ensuremath{\PW^\pm}\xspace}
\def\Z      {\ensuremath{\PZ^0}\xspace}


\def\quark     {\ensuremath{\Pq}\xspace}
\def\quarkbar  {\ensuremath{\overline \quark}\xspace}
\def\qqbar     {\ensuremath{\quark\quarkbar}\xspace}
\def\uquark    {\ensuremath{\Pu}\xspace}
\def\uquarkbar {\ensuremath{\overline \uquark}\xspace}
\def\uubar     {\ensuremath{\uquark\uquarkbar}\xspace}
\def\dquark    {\ensuremath{\Pd}\xspace}
\def\dquarkbar {\ensuremath{\overline \dquark}\xspace}
\def\ddbar     {\ensuremath{\dquark\dquarkbar}\xspace}
\def\squark    {\ensuremath{\Ps}\xspace}
\def\squarkbar {\ensuremath{\overline \squark}\xspace}
\def\ssbar     {\ensuremath{\squark\squarkbar}\xspace}
\def\cquark    {\ensuremath{\Pc}\xspace}
\def\cquarkbar {\ensuremath{\overline \cquark}\xspace}
\def\ccbar     {\ensuremath{\cquark\cquarkbar}\xspace}
\def\bquark    {\ensuremath{\Pb}\xspace}
\def\bquarkbar {\ensuremath{\overline \bquark}\xspace}
\def\bbbar     {\ensuremath{\bquark\bquarkbar}\xspace}
\def\tquark    {\ensuremath{\Pt}\xspace}
\def\tquarkbar {\ensuremath{\overline \tquark}\xspace}
\def\ttbar     {\ensuremath{\tquark\tquarkbar}\xspace}


\def\pion  {\ensuremath{\Ppi}\xspace}
\def\piz   {\ensuremath{\pion^0}\xspace}
\def\pizs  {\ensuremath{\pion^0\mbox\,\rm{s}}\xspace}
\def\ppz   {\ensuremath{\pion^0\pion^0}\xspace}
\def\pip   {\ensuremath{\pion^+}\xspace}
\def\pim   {\ensuremath{\pion^-}\xspace}
\def\pipi  {\ensuremath{\pion^+\pion^-}\xspace}
\def\pipm  {\ensuremath{\pion^\pm}\xspace}
\def\pimp  {\ensuremath{\pion^\mp}\xspace}

\def\kaon  {\ensuremath{\PK}\xspace}
  \def\Kbar  {\kern 0.2em\overline{\kern -0.2em \PK}{}\xspace}
\def\Kb    {\ensuremath{\Kbar}\xspace}
\def\Kz    {\ensuremath{\kaon^0}\xspace}
\def\Kzb   {\ensuremath{\Kbar^0}\xspace}
\def\KzKzb {\ensuremath{\Kz \kern -0.16em \Kzb}\xspace}
\def\Kp    {\ensuremath{\kaon^+}\xspace}
\def\Km    {\ensuremath{\kaon^-}\xspace}
\def\Kpm   {\ensuremath{\kaon^\pm}\xspace}
\def\Kmp   {\ensuremath{\kaon^\mp}\xspace}
\def\KpKm  {\ensuremath{\Kp \kern -0.16em \Km}\xspace}
\def\KS    {\ensuremath{\kaon^0_{\rm\scriptscriptstyle S}}\xspace} 
\def\KSb    {\ensuremath{\Kbar^0_{\rm\scriptscriptstyle S}}\xspace} 
\def\KL    {\ensuremath{\kaon^0_{\rm\scriptscriptstyle L}}\xspace} 
\def\Kstarz  {\ensuremath{\kaon^{*0}}\xspace}
\def\Kstarzb {\ensuremath{\Kbar^{*0}}\xspace}
\def\Kstar   {\ensuremath{\kaon^*}\xspace}
\def\Kstarb  {\ensuremath{\Kbar^*}\xspace}
\def\Kstarp  {\ensuremath{\kaon^{*+}}\xspace}
\def\Kstarm  {\ensuremath{\kaon^{*-}}\xspace}
\def\Kstarpm {\ensuremath{\kaon^{*\pm}}\xspace}
\def\Kstarmp {\ensuremath{\kaon^{*\mp}}\xspace}

\newcommand{\etapr}{\ensuremath{\Peta^{\prime}}\xspace}


  \def\Dbar    {\kern 0.2em\overline{\kern -0.2em \PD}{}\xspace}
\def\D       {\ensuremath{\PD}\xspace}
\def\Db      {\ensuremath{\Dbar}\xspace}
\def\Dz      {\ensuremath{\D^0}\xspace}
\def\Dzb     {\ensuremath{\Dbar^0}\xspace}
\def\DzDzb   {\ensuremath{\Dz {\kern -0.16em \Dzb}}\xspace}
\def\Dp      {\ensuremath{\D^+}\xspace}
\def\Dm      {\ensuremath{\D^-}\xspace}
\def\Dpm     {\ensuremath{\D^\pm}\xspace}
\def\Dmp     {\ensuremath{\D^\mp}\xspace}
\def\DpDm    {\ensuremath{\Dp {\kern -0.16em \Dm}}\xspace}
\def\Dstar   {\ensuremath{\D^*}\xspace}
\def\Dstarb  {\ensuremath{\Dbar^*}\xspace}
\def\Dstarz  {\ensuremath{\D^{*0}}\xspace}
\def\Dstarzb {\ensuremath{\Dbar^{*0}}\xspace}
\def\Dstarp  {\ensuremath{\D^{*+}}\xspace}
\def\Dstarm  {\ensuremath{\D^{*-}}\xspace}
\def\Dstarpm {\ensuremath{\D^{*\pm}}\xspace}
\def\Dstarmp {\ensuremath{\D^{*\mp}}\xspace}
\def\Ds      {\ensuremath{\D^+_\squark}\xspace}
\def\Dsp     {\ensuremath{\D^+_\squark}\xspace}
\def\Dsm     {\ensuremath{\D^-_\squark}\xspace}
\def\Dspm    {\ensuremath{\D^{\pm}_\squark}\xspace}
\def\Dss     {\ensuremath{\D^{*+}_\squark}\xspace}
\def\Dssp    {\ensuremath{\D^{*+}_\squark}\xspace}
\def\Dssm    {\ensuremath{\D^{*-}_\squark}\xspace}
\def\Dsspm   {\ensuremath{\D^{*\pm}_\squark}\xspace}

\def\B       {\ensuremath{\PB}\xspace}
  \def\Bbar    {\kern 0.18em\overline{\kern -0.18em \PB}{}\xspace}
\def\Bb      {\ensuremath{\Bbar}\xspace}
\def\BBbar   {\ensuremath{\B\Bbar}\xspace} 
\def\Bz      {\ensuremath{\B^0}\xspace}
\def\Bzb     {\ensuremath{\Bbar^0}\xspace}
\def\Bu      {\ensuremath{\B^+}\xspace}
\def\Bub     {\ensuremath{\B^-}\xspace}
\def\Bp      {\ensuremath{\Bu}\xspace}
\def\Bm      {\ensuremath{\Bub}\xspace}
\def\Bpm     {\ensuremath{\B^\pm}\xspace}
\def\Bmp     {\ensuremath{\B^\mp}\xspace}
\def\Bd      {\ensuremath{\B^0}\xspace}
\def\Bs      {\ensuremath{\B^0_\squark}\xspace}
\def\Bsb     {\ensuremath{\Bbar^0_\squark}\xspace}
\def\Bstar   {\ensuremath{B_s^*}\xspace}
\def\Bstarb  {\ensuremath{\Bb_s^*}\xspace}
\def\Bdb     {\ensuremath{\Bbar^0}\xspace}
\def\Bc      {\ensuremath{\B_\cquark^+}\xspace}
\def\Bcp     {\ensuremath{\B_\cquark^+}\xspace}
\def\Bcm     {\ensuremath{\B_\cquark^-}\xspace}
\def\Bcpm    {\ensuremath{\B_\cquark^\pm}\xspace}


\def\jpsi     {\ensuremath{{\PJ\mskip -3mu/\mskip -2mu\Ppsi\mskip 2mu}}\xspace}
\def\psitwos  {\ensuremath{\Ppsi{(2S)}}\xspace}
\def\psiprpr  {\ensuremath{\Ppsi(3770)}\xspace}
\def\etac     {\ensuremath{\Peta_\cquark}\xspace}
\def\chiczero {\ensuremath{\Pchi_{\cquark 0}}\xspace}
\def\chicone  {\ensuremath{\Pchi_{\cquark 1}}\xspace}
\def\chictwo  {\ensuremath{\Pchi_{\cquark 2}}\xspace}
  \def\Y#1S{\ensuremath{\PUpsilon{(#1S)}}\xspace}
\def\OneS  {\Y1S}
\def\TwoS  {\Y2S}
\def\ThreeS{\Y3S}
\def\FourS {\Y4S}
\def\FiveS {\Y5S}

\def\chic  {\ensuremath{\Pchi_{c}}\xspace}


\def\proton      {\ensuremath{\Pp}\xspace}
\def\antiproton  {\ensuremath{\overline \proton}\xspace}
\def\neutron     {\ensuremath{\Pn}\xspace}
\def\antineutron {\ensuremath{\overline \neutron}\xspace}

\def\Deltares {\ensuremath{\PDelta}\xspace}
\def\Deltaresbar{\ensuremath{\overline \Deltares}\xspace}
\def\Xires {\ensuremath{\PXi}\xspace}
\def\Xiresbar{\ensuremath{\overline \Xires}\xspace}
\def\L {\ensuremath{\PLambda}\xspace}
\def\Lbar {\ensuremath{\kern 0.1em\overline{\kern -0.1em\Lambda\kern -0.05em}\kern 0.05em{}}\xspace}
\def\Lambdares {\ensuremath{\PLambda}\xspace}
\def\Lambdaresbar{\ensuremath{\Lbar}\xspace}
\def\Sigmares {\ensuremath{\PSigma}\xspace}
\def\Sigmaresbar{\ensuremath{\overline \Sigmares}\xspace}
\def\Omegares {\ensuremath{\POmega}\xspace}
\def\Omegaresbar{\ensuremath{\overline \Omegares}\xspace}


\def\Lb      {\ensuremath{\L^0_\bquark}\xspace}
\def\Lbbar   {\ensuremath{\Lbar^0_\bquark}\xspace}
\def\Lc      {\ensuremath{\L^+_\cquark}\xspace}
\def\Lcbar   {\ensuremath{\Lbar^-_\cquark}\xspace}


\def\BF         {{\ensuremath{\cal B}\xspace}}
\def\BRvis      {{\ensuremath{\BR_{\rm{vis}}}}}
\def\BR         {\BF}
\newcommand{\decay}[2]{\ensuremath{#1\!\to #2}\xspace}         
\def\ra                 {\ensuremath{\rightarrow}\xspace}
\def\to                 {\ensuremath{\rightarrow}\xspace}

\newcommand{\tauBs}{\ensuremath{\tau_{\Bs}}\xspace}
\newcommand{\tauBd}{\ensuremath{\tau_{\Bd}}\xspace}
\newcommand{\tauBz}{\ensuremath{\tau_{\Bz}}\xspace}
\newcommand{\tauBu}{\ensuremath{\tau_{\Bp}}\xspace}
\newcommand{\tauDp}{\ensuremath{\tau_{\Dp}}\xspace}
\newcommand{\tauDz}{\ensuremath{\tau_{\Dz}}\xspace}
\newcommand{\tauL}{\ensuremath{\tau_{\rm L}}\xspace}
\newcommand{\tauH}{\ensuremath{\tau_{\rm H}}\xspace}

\newcommand{\mBd}{\ensuremath{m_{\Bd}}\xspace}
\newcommand{\mBp}{\ensuremath{m_{\Bp}}\xspace}
\newcommand{\mBs}{\ensuremath{m_{\Bs}}\xspace}
\newcommand{\mBc}{\ensuremath{m_{\Bc}}\xspace}
\newcommand{\mLb}{\ensuremath{m_{\Lb}}\xspace}

\def\grpsuthree {\ensuremath{\mathrm{SU}(3)}\xspace}
\def\grpsutw    {\ensuremath{\mathrm{SU}(2)}\xspace}
\def\grpuone    {\ensuremath{\mathrm{U}(1)}\xspace}

\def\ssqtw {\ensuremath{\sin^{2}\!\theta_{\mathrm{W}}}\xspace}
\def\csqtw {\ensuremath{\cos^{2}\!\theta_{\mathrm{W}}}\xspace}
\def\stw   {\ensuremath{\sin\theta_{\mathrm{W}}}\xspace}
\def\ctw   {\ensuremath{\cos\theta_{\mathrm{W}}}\xspace}
\def\ssqtwef {\ensuremath{{\sin}^{2}\theta_{\mathrm{W}}^{\mathrm{eff}}}\xspace}
\def\csqtwef {\ensuremath{{\cos}^{2}\theta_{\mathrm{W}}^{\mathrm{eff}}}\xspace}
\def\stwef {\ensuremath{\sin\theta_{\mathrm{W}}^{\mathrm{eff}}}\xspace}
\def\ctwef {\ensuremath{\cos\theta_{\mathrm{W}}^{\mathrm{eff}}}\xspace}
\def\gv    {\ensuremath{g_{\mbox{\tiny V}}}\xspace}
\def\ga    {\ensuremath{g_{\mbox{\tiny A}}}\xspace}

\def\order   {\ensuremath{\mathcal{O}}\xspace}
\def\ordalph {\ensuremath{\mathcal{O}(\alpha)}\xspace}
\def\ordalsq {\ensuremath{\mathcal{O}(\alpha^{2})}\xspace}
\def\ordalcb {\ensuremath{\mathcal{O}(\alpha^{3})}\xspace}

\newcommand{\as}{\ensuremath{\alpha_{\scriptscriptstyle S}}\xspace}
\newcommand{\MSb}{\ensuremath{\overline{\mathrm{MS}}}\xspace}
\newcommand{\lqcd}{\ensuremath{\Lambda_{\mathrm{QCD}}}\xspace}
\def\qsq       {\ensuremath{q^2}\xspace}


\def\eps   {\ensuremath{\varepsilon}\xspace}
\def\epsK  {\ensuremath{\varepsilon_K}\xspace}
\def\epsB  {\ensuremath{\varepsilon_B}\xspace}
\def\epsp  {\ensuremath{\varepsilon^\prime_K}\xspace}

\def\CP                {\ensuremath{C\!P}\xspace}
\def\CPT               {\ensuremath{C\!PT}\xspace}

\def\rhobar {\ensuremath{\overline \rho}\xspace}
\def\etabar {\ensuremath{\overline \eta}\xspace}

\def\Vud  {\ensuremath{|V_{\uquark\dquark}|}\xspace}
\def\Vcd  {\ensuremath{|V_{\cquark\dquark}|}\xspace}
\def\Vtd  {\ensuremath{|V_{\tquark\dquark}|}\xspace}
\def\Vus  {\ensuremath{|V_{\uquark\squark}|}\xspace}
\def\Vcs  {\ensuremath{|V_{\cquark\squark}|}\xspace}
\def\Vts  {\ensuremath{|V_{\tquark\squark}|}\xspace}
\def\Vub  {\ensuremath{|V_{\uquark\bquark}|}\xspace}
\def\Vcb  {\ensuremath{|V_{\cquark\bquark}|}\xspace}
\def\Vtb  {\ensuremath{|V_{\tquark\bquark}|}\xspace}


\newcommand{\dm}{\ensuremath{\Delta m}\xspace}
\newcommand{\dms}{\ensuremath{\Delta m_{\squark}}\xspace}
\newcommand{\dmd}{\ensuremath{\Delta m_{\dquark}}\xspace}
\newcommand{\DG}{\ensuremath{\Delta\Gamma}\xspace}
\newcommand{\DGs}{\ensuremath{\Delta\Gamma_{\squark}}\xspace}
\newcommand{\DGd}{\ensuremath{\Delta\Gamma_{\dquark}}\xspace}
\newcommand{\Gs}{\ensuremath{\Gamma_{\squark}}\xspace}
\newcommand{\Gd}{\ensuremath{\Gamma_{\dquark}}\xspace}

\newcommand{\MBq}{\ensuremath{M_{\B_\quark}}\xspace}
\newcommand{\DGq}{\ensuremath{\Delta\Gamma_{\quark}}\xspace}
\newcommand{\Gq}{\ensuremath{\Gamma_{\quark}}\xspace}
\newcommand{\dmq}{\ensuremath{\Delta m_{\quark}}\xspace}
\newcommand{\GL}{\ensuremath{\Gamma_{\rm L}}\xspace}
\newcommand{\GH}{\ensuremath{\Gamma_{\rm H}}\xspace}

\newcommand{\DGsGs}{\ensuremath{\Delta\Gamma_{\squark}/\Gamma_{\squark}}\xspace}
\newcommand{\Delm}{\mbox{$\Delta m $}\xspace}
\newcommand{\ACP}{\ensuremath{{\cal A}^{\CP}}\xspace}
\newcommand{\Adir}{\ensuremath{{\cal A}^{\rm dir}}\xspace}
\newcommand{\Amix}{\ensuremath{{\cal A}^{\rm mix}}\xspace}
\newcommand{\ADelta}{\ensuremath{{\cal A}^\Delta}\xspace}
\newcommand{\phid}{\ensuremath{\phi_{\dquark}}\xspace}
\newcommand{\sinphid}{\ensuremath{\sin\!\phid}\xspace}
\newcommand{\phis}{\ensuremath{\phi_{\squark}}\xspace}
\newcommand{\betas}{\ensuremath{\beta_{\squark}}\xspace}
\newcommand{\sbetas}{\ensuremath{\sigma(\beta_{\squark})}\xspace}
\newcommand{\stbetas}{\ensuremath{\sigma(2\beta_{\squark})}\xspace}
\newcommand{\stphis}{\ensuremath{\sigma(\phi_{\squark})}\xspace}
\newcommand{\sinphis}{\ensuremath{\sin\!\phis}\xspace}

\newcommand{\edet}{{\ensuremath{\varepsilon_{\rm det}}}\xspace}
\newcommand{\erec}{{\ensuremath{\varepsilon_{\rm rec/det}}}\xspace}
\newcommand{\esel}{{\ensuremath{\varepsilon_{\rm sel/rec}}}\xspace}
\newcommand{\etrg}{{\ensuremath{\varepsilon_{\rm trg/sel}}}\xspace}
\newcommand{\etot}{{\ensuremath{\varepsilon_{\rm tot}}}\xspace}

\newcommand{\mistag}{\ensuremath{\omega}\xspace}
\newcommand{\wcomb}{\ensuremath{\omega^{\rm comb}}\xspace}
\newcommand{\etag}{{\ensuremath{\varepsilon_{\rm tag}}}\xspace}
\newcommand{\etagcomb}{{\ensuremath{\varepsilon_{\rm tag}^{\rm comb}}}\xspace}
\newcommand{\effeff}{\ensuremath{\varepsilon_{\rm eff}}\xspace}
\newcommand{\effeffcomb}{\ensuremath{\varepsilon_{\rm eff}^{\rm comb}}\xspace}
\newcommand{\efftag}{{\ensuremath{\etag(1-2\omega)^2}}\xspace}
\newcommand{\effD}{{\ensuremath{\etag D^2}}\xspace}

\newcommand{\etagprompt}{{\ensuremath{\varepsilon_{\rm tag}^{\rm Pr}}}\xspace}
\newcommand{\etagLL}{{\ensuremath{\varepsilon_{\rm tag}^{\rm LL}}}\xspace}


\def\BdToKstmm    {\decay{\Bd}{\Kstarz\mup\mun}}
\def\BdbToKstmm   {\decay{\Bdb}{\Kstarzb\mup\mun}}

\def\BsToJPsiPhi  {\decay{\Bs}{\jpsi\phi}}
\def\BdToJPsiKst  {\decay{\Bd}{\jpsi\Kstarz}}
\def\BdbToJPsiKst {\decay{\Bdb}{\jpsi\Kstarzb}}

\def\BsPhiGam     {\decay{\Bs}{\phi \g}}
\def\BdKstGam     {\decay{\Bd}{\Kstarz \g}}

\def\BTohh        {\decay{\B}{\Ph^+ \Ph'^-}}
\def\BdTopipi     {\decay{\Bd}{\pip\pim}}
\def\BdToKpi      {\decay{\Bd}{\Kp\pim}}
\def\BsToKK       {\decay{\Bs}{\Kp\Km}}
\def\BsTopiK      {\decay{\Bs}{\pip\Km}}

\def\BdKstee  {\decay{\Bd}{\Kstarz\epem}}
\def\BdbKstee {\decay{\Bdb}{\Kstarzb\epem}}
\def\bsll     {\decay{\bquark}{\squark \ell^+ \ell^-}}
\def\AFB      {\ensuremath{A_{\mathrm{FB}}}\xspace}
\def\FL       {\ensuremath{F_{\mathrm{L}}}\xspace}
\def\AT#1     {\ensuremath{A_{\mathrm{T}}^{#1}}\xspace}           
\def\btosgam  {\decay{\bquark}{\squark \g}}
\def\btodgam  {\decay{\bquark}{\dquark \g}}
\def\Bsmm     {\decay{\Bs}{\mup\mun}}
\def\Bdmm     {\decay{\Bd}{\mup\mun}}
\def\ctl       {\ensuremath{\cos{\theta_l}}\xspace}
\def\ctk       {\ensuremath{\cos{\theta_K}}\xspace}

\def\C#1      {\ensuremath{\mathcal{C}_{#1}}\xspace}                       
\def\Cp#1     {\ensuremath{\mathcal{C}_{#1}^{'}}\xspace}                    
\def\Ceff#1   {\ensuremath{\mathcal{C}_{#1}^{\mathrm{(eff)}}}\xspace}        
\def\Cpeff#1  {\ensuremath{\mathcal{C}_{#1}^{'\mathrm{(eff)}}}\xspace}       
\def\Ope#1    {\ensuremath{\mathcal{O}_{#1}}\xspace}                       
\def\Opep#1   {\ensuremath{\mathcal{O}_{#1}^{'}}\xspace}                    


\def\xprime     {\ensuremath{x^{\prime}}\xspace}
\def\yprime     {\ensuremath{y^{\prime}}\xspace}
\def\ycp        {\ensuremath{y_{\CP}}\xspace}
\def\agamma     {\ensuremath{A_{\Gamma}}\xspace}
\def\kpi        {\ensuremath{\PK\Ppi}\xspace}
\def\kk         {\ensuremath{\PK\PK}\xspace}
\def\dkpi       {\decay{\PD}{\PK\Ppi}}
\def\dkk        {\decay{\PD}{\PK\PK}}
\def\dkpicf     {\decay{\Dz}{\Km\pip}}

\newcommand{\bra}[1]{\ensuremath{\langle #1|}}             
\newcommand{\ket}[1]{\ensuremath{|#1\rangle}}              
\newcommand{\braket}[2]{\ensuremath{\langle #1|#2\rangle}} 

\newcommand{\unit}[1]{\ensuremath{\rm\,#1}\xspace}          

\newcommand{\tev}{\ensuremath{\mathrm{\,Te\kern -0.1em V}}\xspace}
\newcommand{\gev}{\ensuremath{\mathrm{\,Ge\kern -0.1em V}}\xspace}
\newcommand{\mev}{\ensuremath{\mathrm{\,Me\kern -0.1em V}}\xspace}
\newcommand{\kev}{\ensuremath{\mathrm{\,ke\kern -0.1em V}}\xspace}
\newcommand{\ev}{\ensuremath{\mathrm{\,e\kern -0.1em V}}\xspace}
\newcommand{\gevc}{\ensuremath{{\mathrm{\,Ge\kern -0.1em V\!/}c}}\xspace}
\newcommand{\mevc}{\ensuremath{{\mathrm{\,Me\kern -0.1em V\!/}c}}\xspace}
\newcommand{\gevcc}{\ensuremath{{\mathrm{\,Ge\kern -0.1em V\!/}c^2}}\xspace}
\newcommand{\gevgevcccc}{\ensuremath{{\mathrm{\,Ge\kern -0.1em V^2\!/}c^4}}\xspace}
\newcommand{\mevcc}{\ensuremath{{\mathrm{\,Me\kern -0.1em V\!/}c^2}}\xspace}

\def\km   {\ensuremath{\rm \,km}\xspace}
\def\m    {\ensuremath{\rm \,m}\xspace}
\def\cm   {\ensuremath{\rm \,cm}\xspace}
\def\cma  {\ensuremath{{\rm \,cm}^2}\xspace}
\def\mm   {\ensuremath{\rm \,mm}\xspace}
\def\mma  {\ensuremath{{\rm \,mm}^2}\xspace}
\def\mum  {\ensuremath{\,\upmu\rm m}\xspace}
\def\muma {\ensuremath{\,\upmu\rm m^2}\xspace}
\def\nm   {\ensuremath{\rm \,nm}\xspace}
\def\fm   {\ensuremath{\rm \,fm}\xspace}
\def\barn{\ensuremath{\rm \,b}\xspace}
\def\barnhyph{\ensuremath{\rm -b}\xspace}
\def\mbarn{\ensuremath{\rm \,mb}\xspace}
\def\mub{\ensuremath{\rm \,\upmu b}\xspace}
\def\mbarnhyph{\ensuremath{\rm -mb}\xspace}
\def\nb {\ensuremath{\rm \,nb}\xspace}
\def\invnb {\ensuremath{\mbox{\,nb}^{-1}}\xspace}
\def\pb {\ensuremath{\rm \,pb}\xspace}
\def\invpb {\ensuremath{\mbox{\,pb}^{-1}}\xspace}
\def\fb   {\ensuremath{\mbox{\,fb}}\xspace}
\def\invfb   {\ensuremath{\mbox{\,fb}^{-1}}\xspace}

\def\sec  {\ensuremath{\rm {\,s}}\xspace}
\def\ms   {\ensuremath{{\rm \,ms}}\xspace}
\def\mus  {\ensuremath{\,\upmu{\rm s}}\xspace}
\def\ns   {\ensuremath{{\rm \,ns}}\xspace}
\def\ps   {\ensuremath{{\rm \,ps}}\xspace}
\def\fs   {\ensuremath{\rm \,fs}\xspace}

\def\mhz  {\ensuremath{{\rm \,MHz}}\xspace}
\def\khz  {\ensuremath{{\rm \,kHz}}\xspace}
\def\hz   {\ensuremath{{\rm \,Hz}}\xspace}

\def\invps{\ensuremath{{\rm \,ps^{-1}}}\xspace}

\def\yr   {\ensuremath{\rm \,yr}\xspace}
\def\hr   {\ensuremath{\rm \,hr}\xspace}

\def\degc {\ensuremath{^\circ}{C}\xspace}
\def\degk {\ensuremath {\rm K}\xspace}

\def\Xrad {\ensuremath{X_0}\xspace}
\def\NIL{\ensuremath{\lambda_{int}}\xspace}
\def\mip {MIP\xspace}
\def\neutroneq {\ensuremath{\rm \,n_{eq}}\xspace}
\def\neqcmcm {\ensuremath{\rm \,n_{eq} / cm^2}\xspace}
\def\kRad {\ensuremath{\rm \,kRad}\xspace}
\def\MRad {\ensuremath{\rm \,MRad}\xspace}
\def\ci {\ensuremath{\rm \,Ci}\xspace}
\def\mci {\ensuremath{\rm \,mCi}\xspace}

\def\sx    {\ensuremath{\sigma_x}\xspace}    
\def\sy    {\ensuremath{\sigma_y}\xspace}   
\def\sz    {\ensuremath{\sigma_z}\xspace}    

\newcommand{\stat}{\ensuremath{\mathrm{(stat)}}\xspace}
\newcommand{\syst}{\ensuremath{\mathrm{(syst)}}\xspace}


\def\order{{\ensuremath{\cal O}}\xspace}
\newcommand{\chisq}{\ensuremath{\chi^2}\xspace}

\def\deriv {\ensuremath{\mathrm{d}}}

\def\gsim{{~\raise.15em\hbox{$>$}\kern-.85em
          \lower.35em\hbox{$\sim$}~}\xspace}
\def\lsim{{~\raise.15em\hbox{$<$}\kern-.85em
          \lower.35em\hbox{$\sim$}~}\xspace}

\newcommand{\mean}[1]{\ensuremath{\left\langle #1 \right\rangle}} 
\newcommand{\abs}[1]{\ensuremath{\left\|#1\right\|}} 
\newcommand{\Real}{\ensuremath{\mathcal{R}e}\xspace}
\newcommand{\Imag}{\ensuremath{\mathcal{I}m}\xspace}

\def\PDF {PDF\xspace}

\def\Ebeam {\ensuremath{E_{\mbox{\tiny BEAM}}}\xspace}
\def\sqs   {\ensuremath{\protect\sqrt{s}}\xspace}

\def\ptot       {\mbox{$p$}\xspace}
\def\pt         {\mbox{$p_{\rm T}$}\xspace}
\def\et         {\mbox{$E_{\rm T}$}\xspace}
\def\dpp        {\ensuremath{\mathrm{d}\hspace{-0.1em}p/p}\xspace}

\newcommand{\dedx}{\ensuremath{\mathrm{d}\hspace{-0.1em}E/\mathrm{d}x}\xspace}


\def\dllkpi     {\ensuremath{\mathrm{DLL}_{\kaon\pion}}\xspace}
\def\dllppi     {\ensuremath{\mathrm{DLL}_{\proton\pion}}\xspace}
\def\dllepi     {\ensuremath{\mathrm{DLL}_{\electron\pion}}\xspace}
\def\dllmupi    {\ensuremath{\mathrm{DLL}_{\mmu\pi}}\xspace}

\def\mphi       {\mbox{$\phi$}\xspace}
\def\mtheta     {\mbox{$\theta$}\xspace}
\def\ctheta     {\mbox{$\cos\theta$}\xspace}
\def\stheta     {\mbox{$\sin\theta$}\xspace}
\def\ttheta     {\mbox{$\tan\theta$}\xspace}

\def\degrees{\ensuremath{^{\circ}}\xspace}
\def\krad {\ensuremath{\rm \,krad}\xspace}
\def\mrad{\ensuremath{\rm \,mrad}\xspace}
\def\rad{\ensuremath{\rm \,rad}\xspace}

\def\betastar {\ensuremath{\beta^*}}
\newcommand{\lum} {\ensuremath{\mathcal{L}}\xspace}
\newcommand{\intlum}[1]{\ensuremath{\int\lum=#1\xspace}}  


\def\evtgen     {\mbox{\textsc{EvtGen}}\xspace}
\def\pythia     {\mbox{\textsc{Pythia}}\xspace}
\def\fluka      {\mbox{\textsc{Fluka}}\xspace}
\def\tosca      {\mbox{\textsc{Tosca}}\xspace}
\def\ansys      {\mbox{\textsc{Ansys}}\xspace}
\def\spice      {\mbox{\textsc{Spice}}\xspace}
\def\garfield   {\mbox{\textsc{Garfield}}\xspace}
\def\geant      {\mbox{\textsc{Geant3}}\xspace}
\def\hepmc      {\mbox{\textsc{HepMC}}\xspace}
\def\gauss      {\mbox{\textsc{Gauss}}\xspace}
\def\gaudi      {\mbox{\textsc{Gaudi}}\xspace}
\def\boole      {\mbox{\textsc{Boole}}\xspace}
\def\brunel     {\mbox{\textsc{Brunel}}\xspace}
\def\davinci    {\mbox{\textsc{DaVinci}}\xspace}
\def\erasmus    {\mbox{\textsc{Erasmus}}\xspace}
\def\moore      {\mbox{\textsc{Moore}}\xspace}
\def\ganga      {\mbox{\textsc{Ganga}}\xspace}
\def\dirac      {\mbox{\textsc{Dirac}}\xspace}
\def\root       {\mbox{\textsc{Root}}\xspace}
\def\roofit     {\mbox{\textsc{RooFit}}\xspace}
\def\pyroot     {\mbox{\textsc{PyRoot}}\xspace}
\def\photos     {\mbox{\textsc{Photos}}\xspace}

\def\cpp        {\mbox{\textsc{C\raisebox{0.1em}{{\footnotesize{++}}}}}\xspace}
\def\python     {\mbox{\textsc{Python}}\xspace}
\def\ruby       {\mbox{\textsc{Ruby}}\xspace}
\def\fortran    {\mbox{\textsc{Fortran}}\xspace}
\def\svn        {\mbox{\textsc{SVN}}\xspace}

\def\kbytes     {\ensuremath{{\rm \,kbytes}}\xspace}
\def\kbsps      {\ensuremath{{\rm \,kbytes/s}}\xspace}
\def\kbits      {\ensuremath{{\rm \,kbits}}\xspace}
\def\kbsps      {\ensuremath{{\rm \,kbits/s}}\xspace}
\def\mbsps      {\ensuremath{{\rm \,Mbits/s}}\xspace}
\def\mbytes     {\ensuremath{{\rm \,Mbytes}}\xspace}
\def\mbps       {\ensuremath{{\rm \,Mbyte/s}}\xspace}
\def\mbsps      {\ensuremath{{\rm \,Mbytes/s}}\xspace}
\def\gbsps      {\ensuremath{{\rm \,Gbits/s}}\xspace}
\def\gbytes     {\ensuremath{{\rm \,Gbytes}}\xspace}
\def\gbsps      {\ensuremath{{\rm \,Gbytes/s}}\xspace}
\def\tbytes     {\ensuremath{{\rm \,Tbytes}}\xspace}
\def\tbpy       {\ensuremath{{\rm \,Tbytes/yr}}\xspace}

\def\dst        {DST\xspace}


\def\nonn {\ensuremath{\rm {\it{n^+}}\mbox{-}on\mbox{-}{\it{n}}}\xspace}
\def\ponn {\ensuremath{\rm {\it{p^+}}\mbox{-}on\mbox{-}{\it{n}}}\xspace}
\def\nonp {\ensuremath{\rm {\it{n^+}}\mbox{-}on\mbox{-}{\it{p}}}\xspace}
\def\cvd  {CVD\xspace}
\def\mwpc {MWPC\xspace}
\def\gem  {GEM\xspace}

\def\tell1  {TELL1\xspace}
\def\ukl1   {UKL1\xspace}
\def\beetle {Beetle\xspace}
\def\otis   {OTIS\xspace}
\def\croc   {CROC\xspace}
\def\carioca {CARIOCA\xspace}
\def\dialog {DIALOG\xspace}
\def\sync   {SYNC\xspace}
\def\cardiac {CARDIAC\xspace}
\def\gol    {GOL\xspace}
\def\vcsel  {VCSEL\xspace}
\def\ttc    {TTC\xspace}
\def\ttcrx  {TTCrx\xspace}
\def\hpd    {HPD\xspace}
\def\pmt    {PMT\xspace}
\def\specs  {SPECS\xspace}
\def\elmb   {ELMB\xspace}
\def\fpga   {FPGA\xspace}
\def\plc    {PLC\xspace}
\def\rasnik {RASNIK\xspace}
\def\elmb   {ELMB\xspace}
\def\can    {CAN\xspace}
\def\lvds   {LVDS\xspace}
\def\ntc    {NTC\xspace}
\def\adc    {ADC\xspace}
\def\led    {LED\xspace}
\def\ccd    {CCD\xspace}
\def\hv     {HV\xspace}
\def\lv     {LV\xspace}
\def\pvss   {PVSS\xspace}
\def\cmos   {CMOS\xspace}
\def\fifo   {FIFO\xspace}
\def\ccpc   {CCPC\xspace}

\def\cfourften     {\ensuremath{\rm C_4 F_{10}}\xspace}
\def\cffour        {\ensuremath{\rm CF_4}\xspace}
\def\cotwo         {\ensuremath{\rm CO_2}\xspace} 
\def\csixffouteen  {\ensuremath{\rm C_6 F_{14}}\xspace} 
\def\mgftwo     {\ensuremath{\rm Mg F_2}\xspace} 
\def\siotwo     {\ensuremath{\rm SiO_2}\xspace} 

\newcommand{\eg}{\mbox{\itshape e.g.}\xspace}
\newcommand{\ie}{\mbox{\itshape i.e.}}
\newcommand{\etal}{{\slshape et al.\/}\xspace}
\newcommand{\etc}{\mbox{\itshape etc.}\xspace}
\newcommand{\cf}{\mbox{\itshape cf.}\xspace}
\newcommand{\ffp}{\mbox{\itshape ff.}\xspace}
 
\widetext


\title{Charmless two-body decays at \belle and prospects at \belle II}
\author{Bilas Pal \\ Brookhaven National Laboratory, Upton, NY\\ (On behalf of the Belle and Belle II Collaborations) }


\begin{abstract}
Charmless hadronic $B$ decays are a good testing ground for the Standard Model (SM) of Particle Physics. The dominant amplitudes are CKM suppressed tree diagrams and/or $ b\to s$ or $b\to d$ loop (``penguin") diagrams. Non-SM particle could appear in the loop, and hence these decays are sensitive to search for New Physics  (NP). Some of the recent measurements of two-body charmless hadronic $B$ decays from \belle and their prospects at \belle II are discussed.
\end{abstract}

\maketitle

\section{Introduction}
Charmless hadronic final states in $B$ decays have branching fractions of order $10^{-5}$ or less, since either the final state is reached by the $b \to u$ transition, which is suppressed by the small CKM matrix element $|V_{ub}|$, or the transition is loop-suppressed. Charmless decays are a good place to observe direct \CP violation, since the smallness of the leading amplitude often implies that another amplitude with a different CKM factor is of similar size. If the two amplitudes also have a substantial (strong) phase difference, this leads to size-able direct \CP violation, which has indeed been observed. There is a large number of potentially interesting decay modes. 
In this proceeding, an overview of the recent measurements of  two-body hadronic $B$ decays at \belle and their prospects at \belle II are discussed. 
\section{Evidence for the decay \boldmath{$\Bd\to \eta \piz$}}
The decay $\Bd\to \eta \piz$  proceeds mainly via a $b\to u$ Cabibbo- and
color-suppressed ``tree'' diagram, and via a $b\to d$ 
``penguin" diagram, as shown
in Fig.~\ref{fig:feynman}. 
The branching fraction of this decay mode can
be used to constrain isospin-breaking effects on the value of
$\sin2\phi_2~(\sin 2\alpha)$  measured in $B\to\pi\pi$
decays~\cite{Gronau:2005pq,Gardner:2005pq}.
It can also be used
to constrain  \CP-violating parameters ($\mathcal{C}^{}_{\eta' K}$
and $\mathcal{S}^{}_{\eta' K}$) governing the time dependence of
$\Bd\to\eta' K^0$ decays~\cite{etapKs_bound}.
The branching fraction is estimated 
using QCD factorization~\cite{qcd}, soft collinear effective
field theory~\cite{Williamson:2006hb}, and flavor SU(3) symmetry~\cite{su3}
and is found to be in the range $(2 - 12)\times10^{-7}$.
\begin{figure}[htb]
\begin{center}
\includegraphics[width=0.24\textwidth]{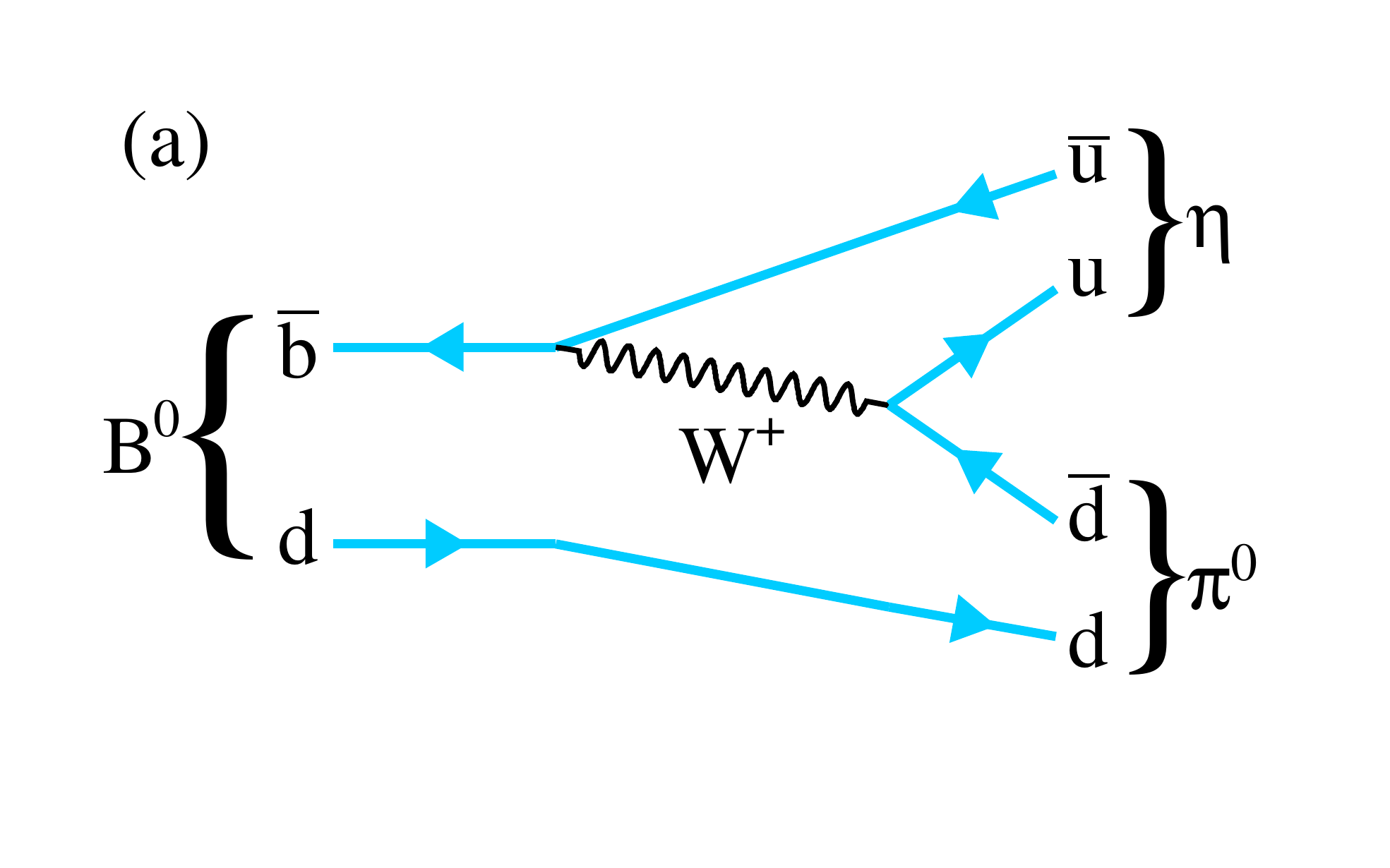}%
\includegraphics[width=0.24\textwidth]{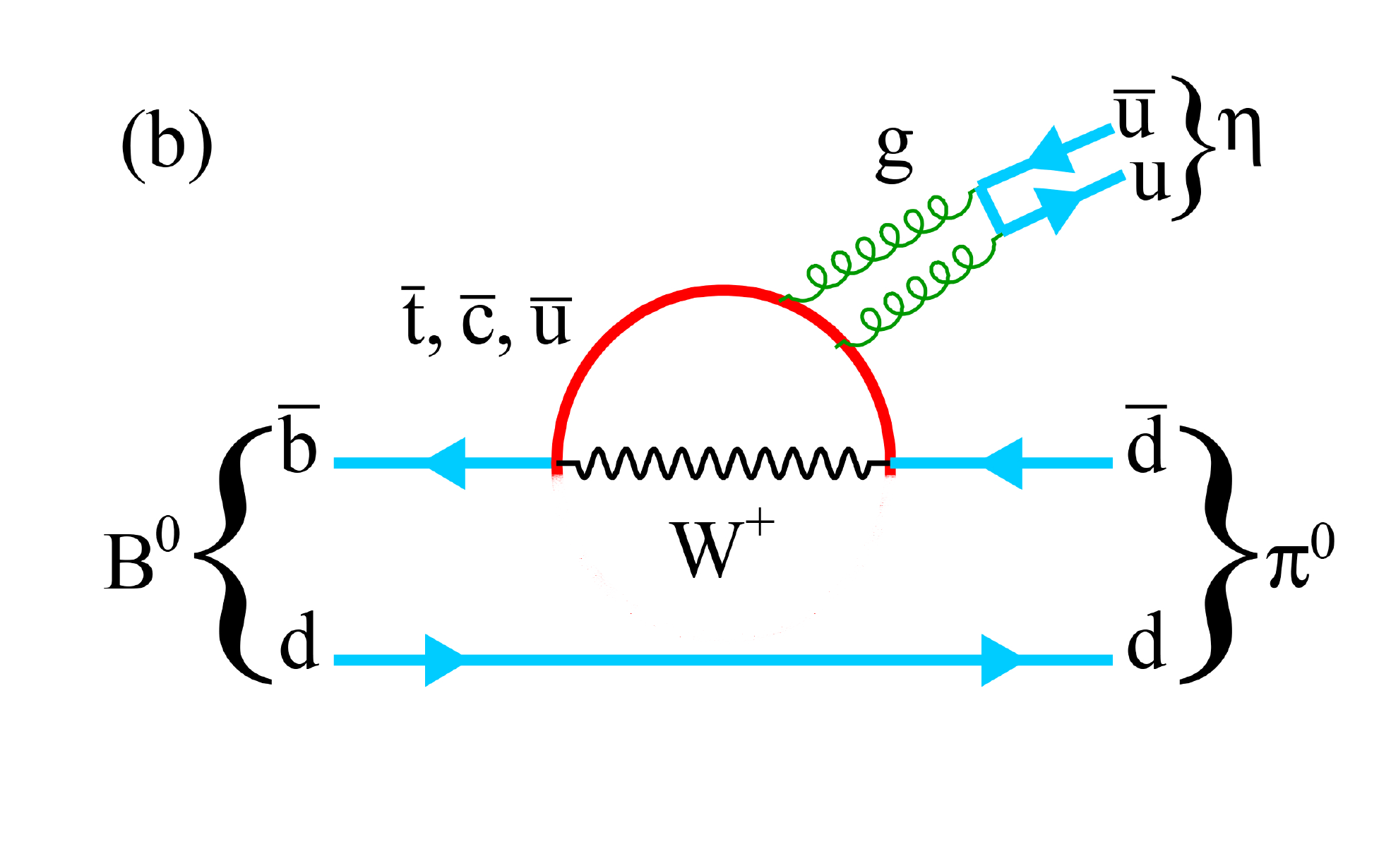}
\vskip -0.75cm
\caption{\small (a) Tree  and (b) penguin diagram 
contributions to  $B\rightarrow \eta \pi^0$ . }
\label{fig:feynman}
\end{center}
\end{figure}

Several experiments~\cite{Albrecht:1990am, Acciarri:1995bx, Richichi:1999kj, Chang:2004fz, Aubert:2008fu}, including \belle, have searched for this decay mode.
The most stringent limit on the branching fraction
is $\mathcal{B}(B\to\eta\piz)<1.5\times10^{-6}$ at 
90\% confidence level (C.L.), given by \babar experiment~\cite{Aubert:2008fu}. The analysis presented here uses  the full data set of the \belle experiment running on
the $\FourS$ resonance at the KEKB asymmetric-energy 
$\epem$ collider. This data set corresponds to
$753\times10^{6}$ $\BBbar$ pairs, which is a factor of 5
larger than that used previously. Improved tracking, photon reconstruction,
and continuum suppression algorithms are also used in this analysis.

\begin{figure}[h!t!p!]
\begin{center}
    \includegraphics[width=0.24\textwidth]{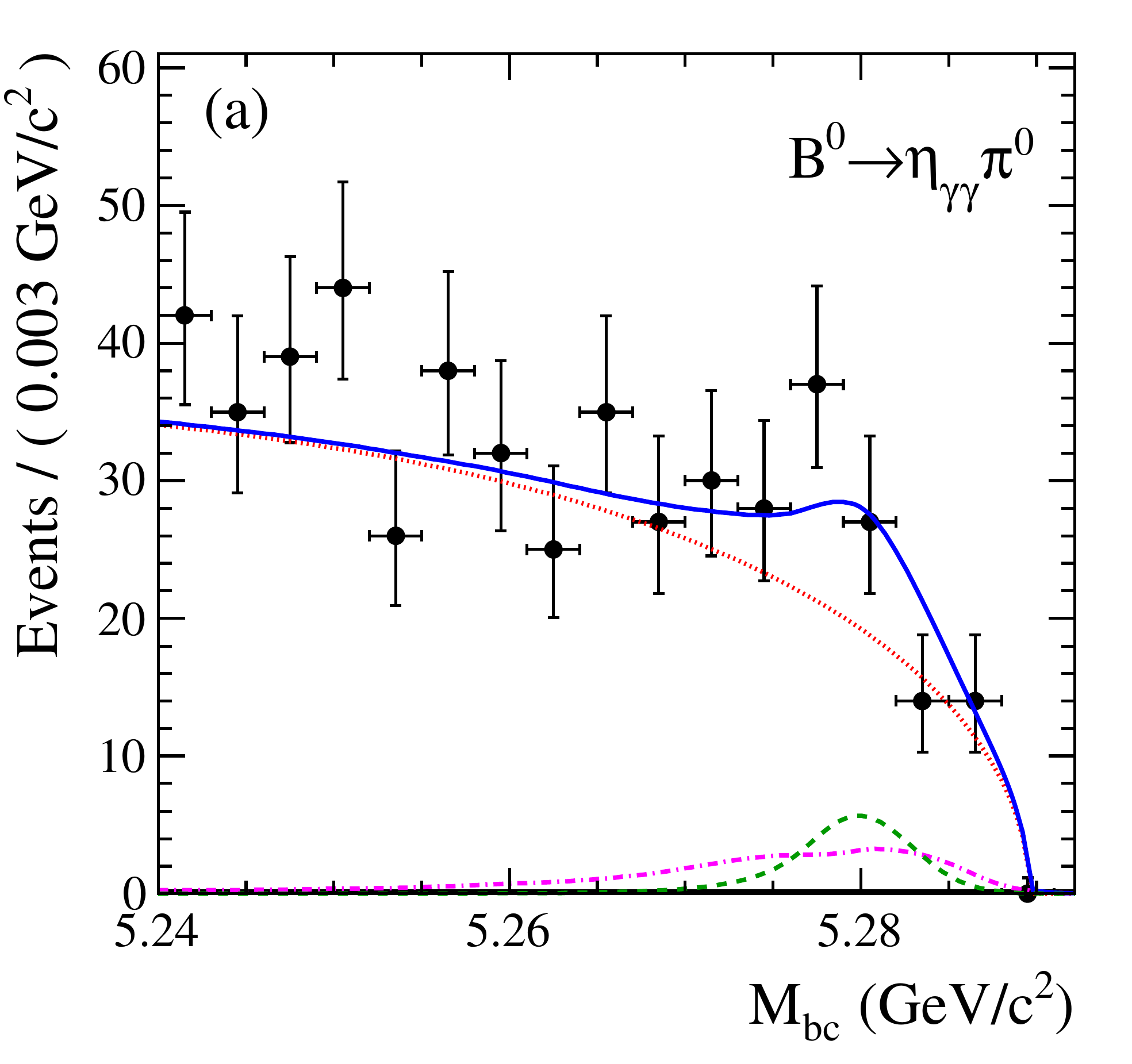}%
    \includegraphics[width=0.24\textwidth]{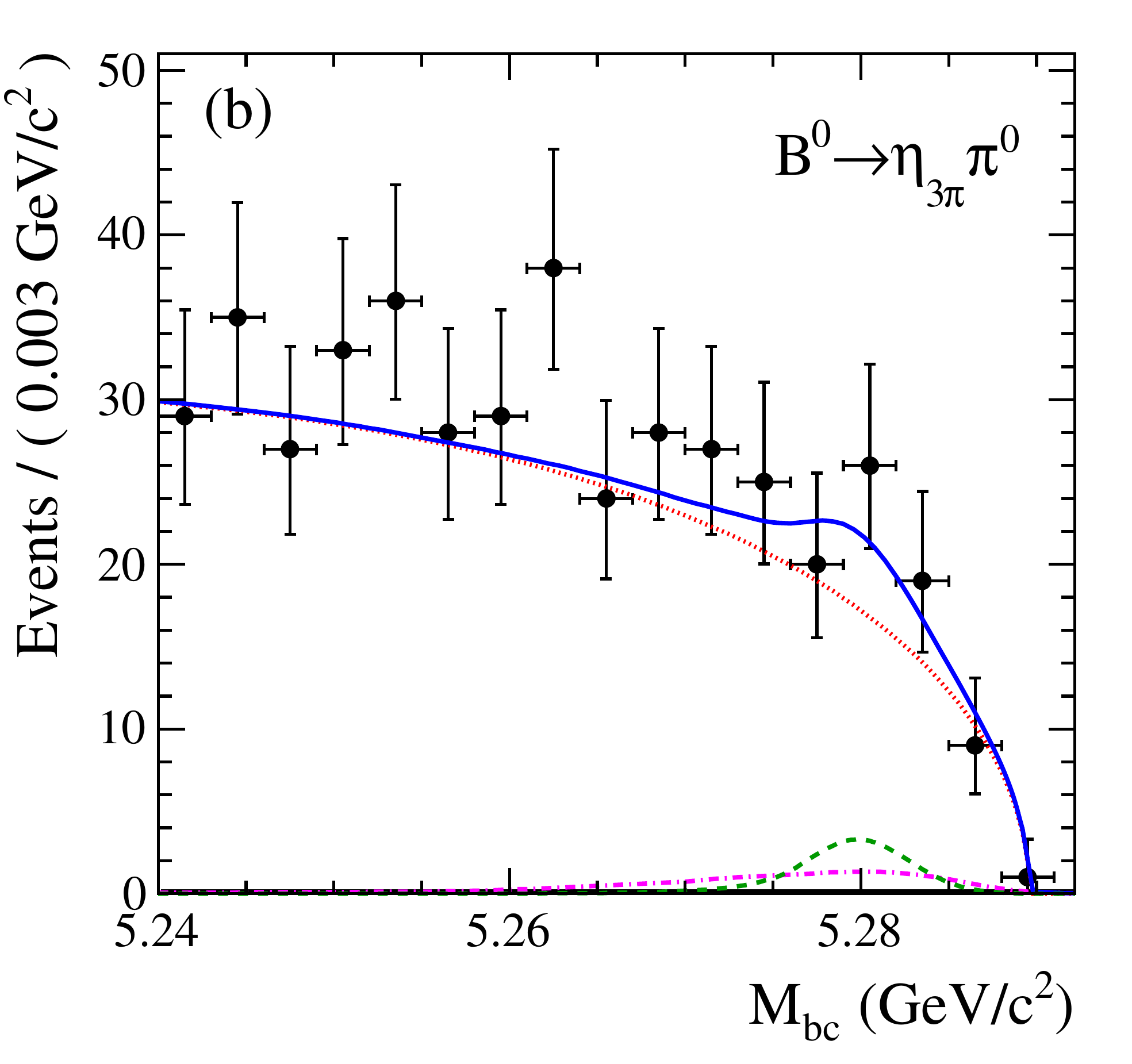}
    \includegraphics[width=0.24\textwidth]{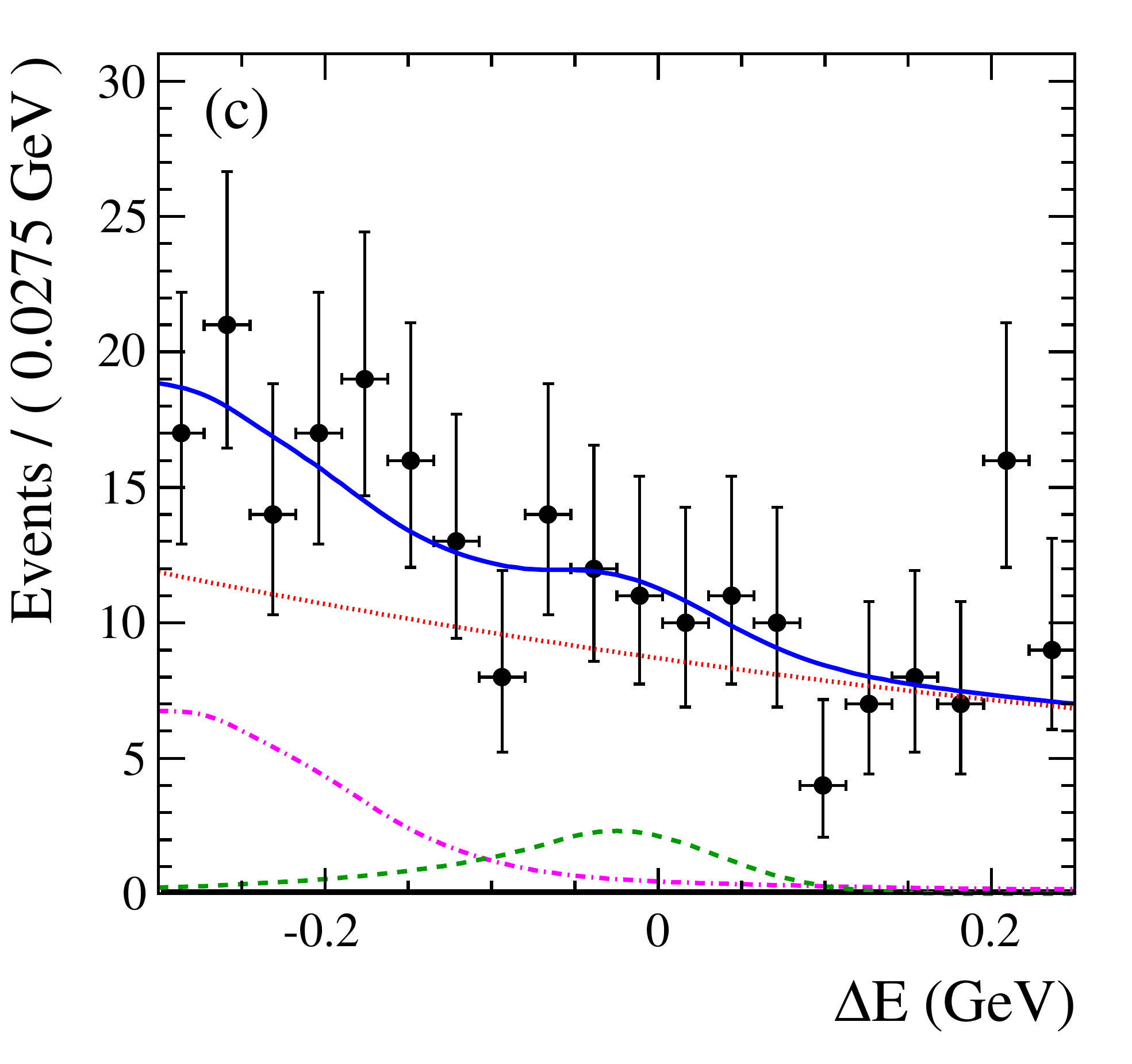}%
     \includegraphics[width=0.24\textwidth]{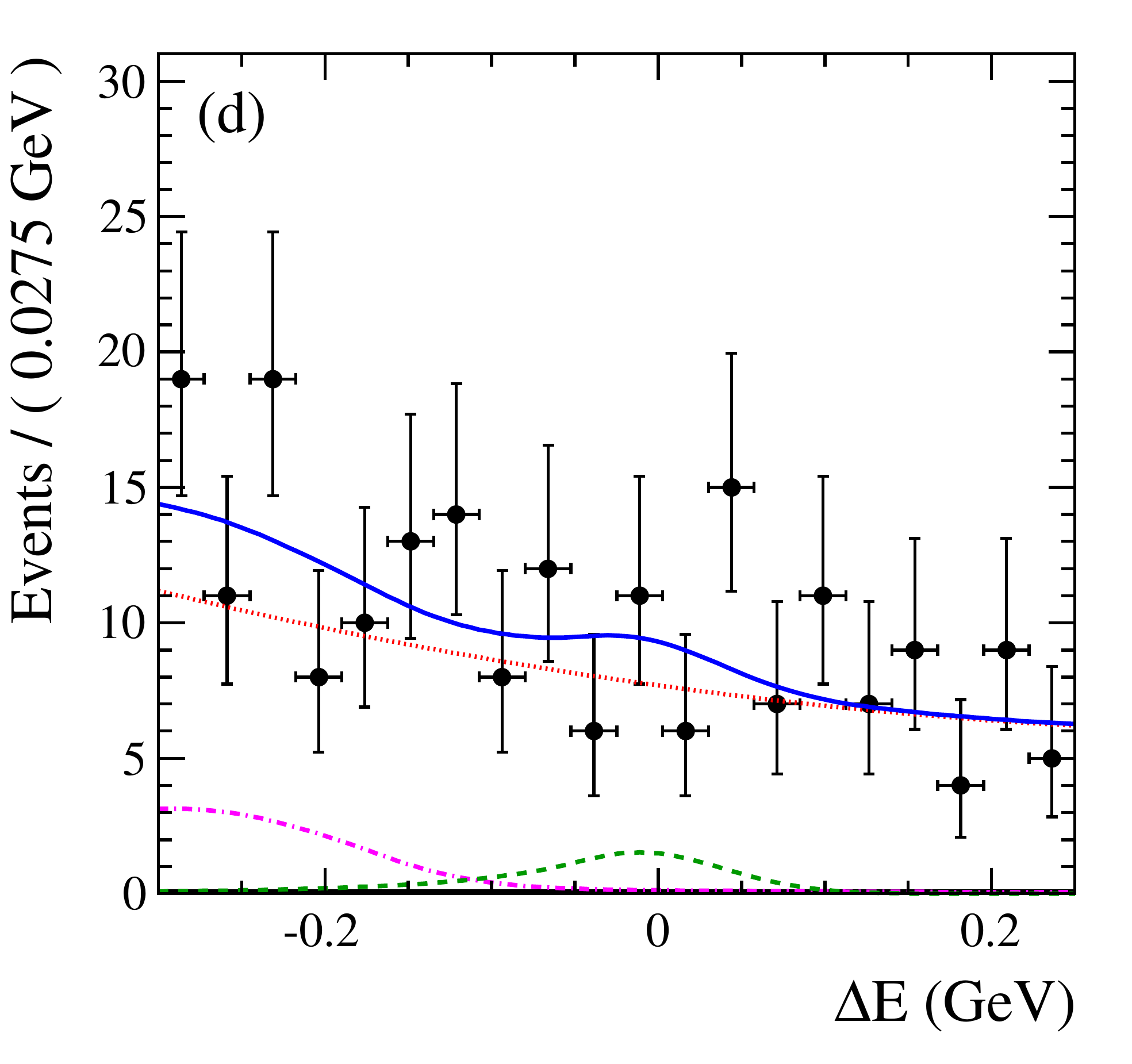}
    \includegraphics[width=0.24\textwidth]{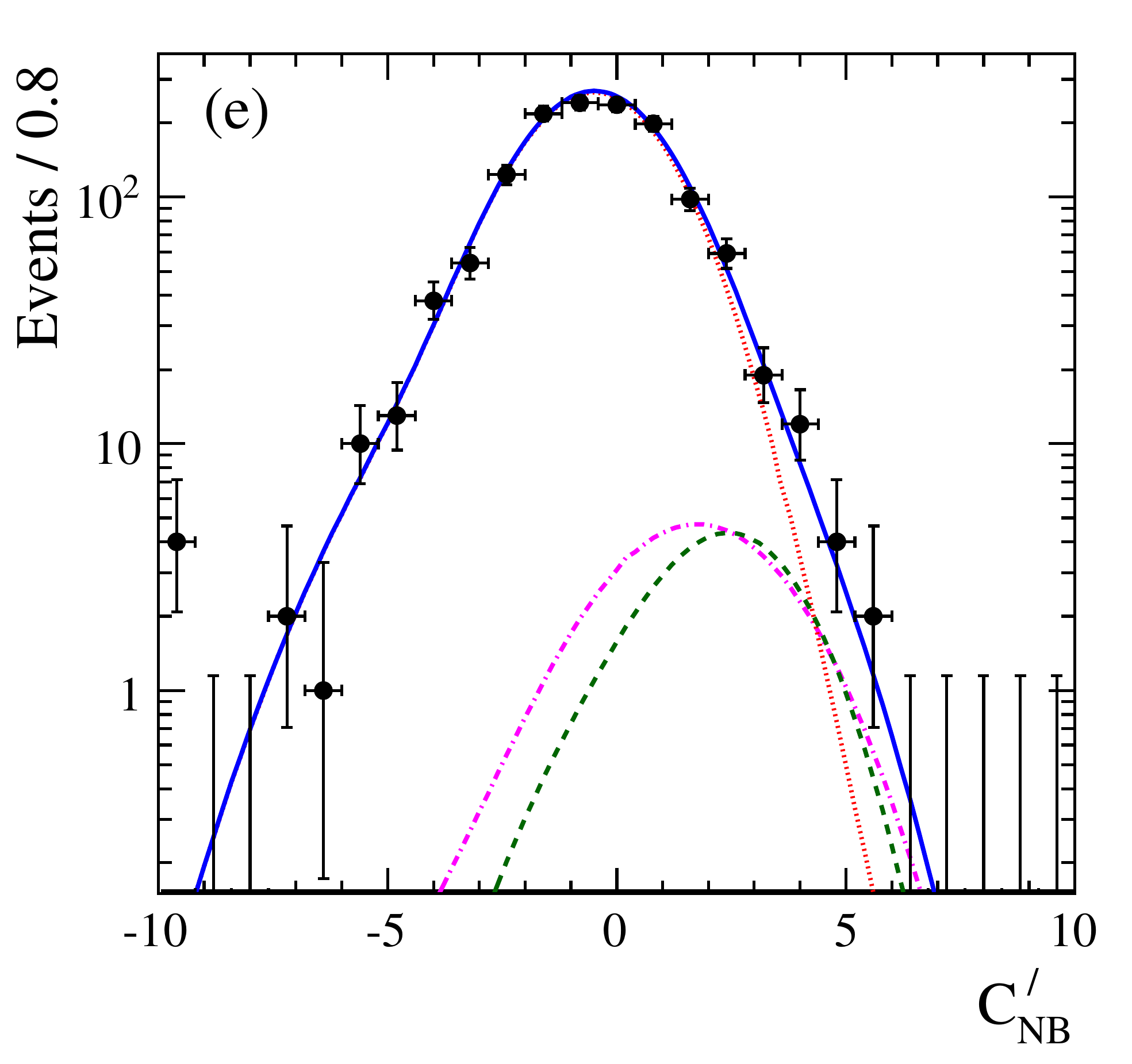}%
    \includegraphics[width=0.24\textwidth]{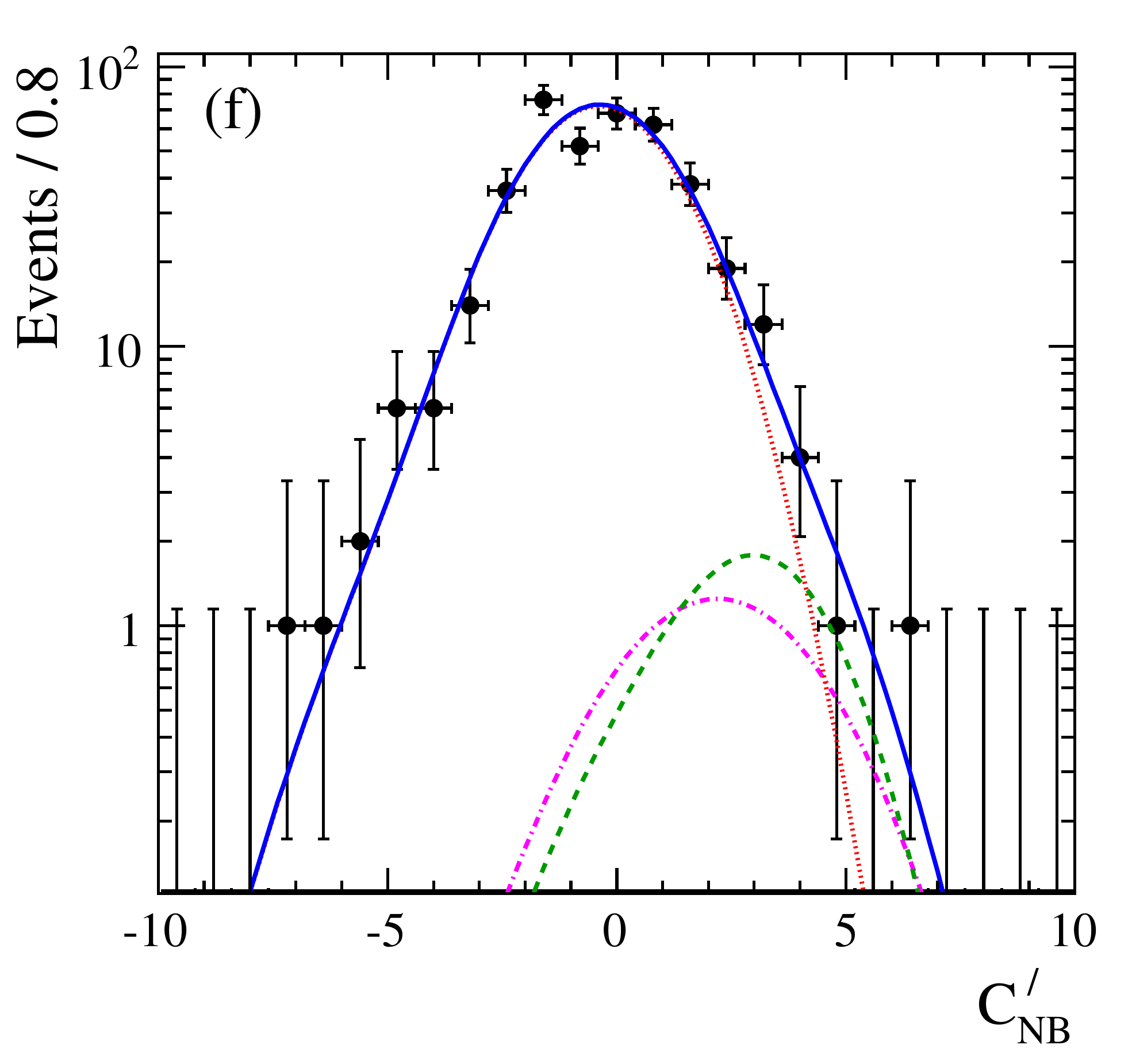}
\end{center}
\vskip -0.5cm
\caption{\small Signal enhanced  projections of the simultaneous fit for the decay $\Bd\to\eta\piz$:
(a), (b) $M_{\rm bc}$; (c), (d) $\Delta E$;
(e), (f) $C'_{\rm NB}$.
The top (bottom) row corresponds to $\eta\to\gamma\gamma$
($\eta\to\pi^+ \pi^-\pi^0$) decays. 
Points with error bars are data;  the (green) dashed, (red) dotted and  (magenta) dot-dashed  curves represent the signal,  continuum and charmless rare  backgrounds, respectively, and the (blue) solid curves represent
the total PDF.}
\label{fig:real_full}
\end{figure}

We find the evidence of the decay $\Bd\to \eta \piz$~\cite{Pal:2015ewa}, where the candidate $\eta$ mesons
are reconstructed via $\eta\to\gamma\gamma~(\eta_{\gamma\gamma})$ and $\eta\to\pip\pim\piz~(\eta_{3\pi})$ decays and 
$\piz$ via $\piz\to\gamma\gamma$. Results of the fit to the variables, beam-energy-constrained mass  $M_{\rm bc}=\sqrt{E^2_{\rm beam}-|\vec{p}_B|^2c^2}/c^2$, energy difference $\Delta E=E_B-E_{\rm beam}$ and  continuum suppression variable $C'_{\rm NB}=\ln(\frac{C_{\rm NB}-C^{\rm min}_{\rm NB}}{C^{\rm max}_{\rm NB}-C_{\rm NB}})$,  are given in Table.~\ref{fit}.
\begin{table}[htb]
\begin{center}
\renewcommand{\arraystretch}{1.5}
\caption{\small Fitted signal yield $Y_{\rm sig}$,
reconstruction efficiency $\epsilon$,
$\eta$ decay branching fraction $\mathcal{B}_{\eta}$,
signal significance, and $B^0$ branching fraction 
$\mathcal{B}$ for the decay $\Bd\to\eta\piz$. The errors listed are statistical only.
The significance  includes both statistical and
systematic uncertainties.}
\label{fit}
\begin{tabular}{c|ccccc}
\hline \hline
Mode & $Y_{\rm sig}$ & $\epsilon(\%)$ & $\mathcal{B}_{\eta}(\%)$ & 
Significance & $\mathcal{B}(10^{-7})$\\
\hline
$B^0\to\eta_{\gamma\gamma}\pi^0$ & $30.6^{+12.2}_{-10.8}$ &
       18.4 & 39.41 & 3.1 & $5.6^{+2.2}_{-2.0}$ \\
$B^0\to\eta_{3\pi}\pi^0$ & $0.5^{+6.6}_{-5.4}$ &
       14.2 & 22.92 & 0.1 & $0.2^{+2.8}_{-2.3}$ \\
Combined & & & & 3.0 & $4.1^{+1.7}_{-1.5}$ \\
\hline\hline
\end{tabular}
\end{center}
\end{table}
The combined branching fraction
is determined by simultaneously fitting both $\Bd\to\eta_{\gamma\gamma}\piz$
and $\Bd\to\eta_{3\pi}\piz$ samples for a common  $\mathcal{B}(\Bd\to\eta\piz)$.  Signal enhanced projections of the simultaneous fit are
shown in Fig.~\ref{fig:real_full}.
The branching
fraction for $\Bd\to \eta \piz$  decays is measured to be
\begin{equation*}
\mathcal{B}(\Bd\to\eta\piz) = 
\left( 4.1^{+1.7+0.5}_{-1.5-0.7}\right) \times 10^{-7}
\end{equation*}
where the first uncertainty is statistical and the second
is systematic. This corresponds to a 90\% C.L. upper limit of $\mathcal{B}(\Bd\to\eta\piz)<6.5\times 10^{-7}$. The significance of this result is~$3.0$ standard deviations, and thus this measurement constitutes the first evidence for this decay.
The measured branching fraction  is in good agreement with theoretical
expectations~\cite{qcd, Williamson:2006hb, su3}. Inserting our measured value
into Eq. (19) of Ref.~\cite{Gronau:2005pq} gives the result that the
isospin-breaking correction  to the  weak phase $\phi_2$ measured in $\B\to\pi\pi$ decays due to $\piz$--$\eta$--$\eta'$ mixing is less than $0.97^{\circ}$ at 90\% C.L.
\section{Evidence for the decay \boldmath{$\Bd\to \eta \eta$}}
This decay is similar to the decay discussed in the previous section.
Previously this decay mode is studied by the \belle, \babar, \cleo and L3 experiments~\cite{PDG} and the most 
stringent limit on the branching fraction [$\mathcal{B}(B\to\eta\eta)<1.0\times10^{-6}$ at 90\% C.L.] is set by the \babar experiment~\cite{Aubert:2009yx}.
Here we update the previous \belle result by using the full data set of the \belle experiment running on the 
\FourS~\cite{Abdesselam:2016tpr}.

\begin{figure}[htb]
\centering
\includegraphics[width=0.5\textwidth]{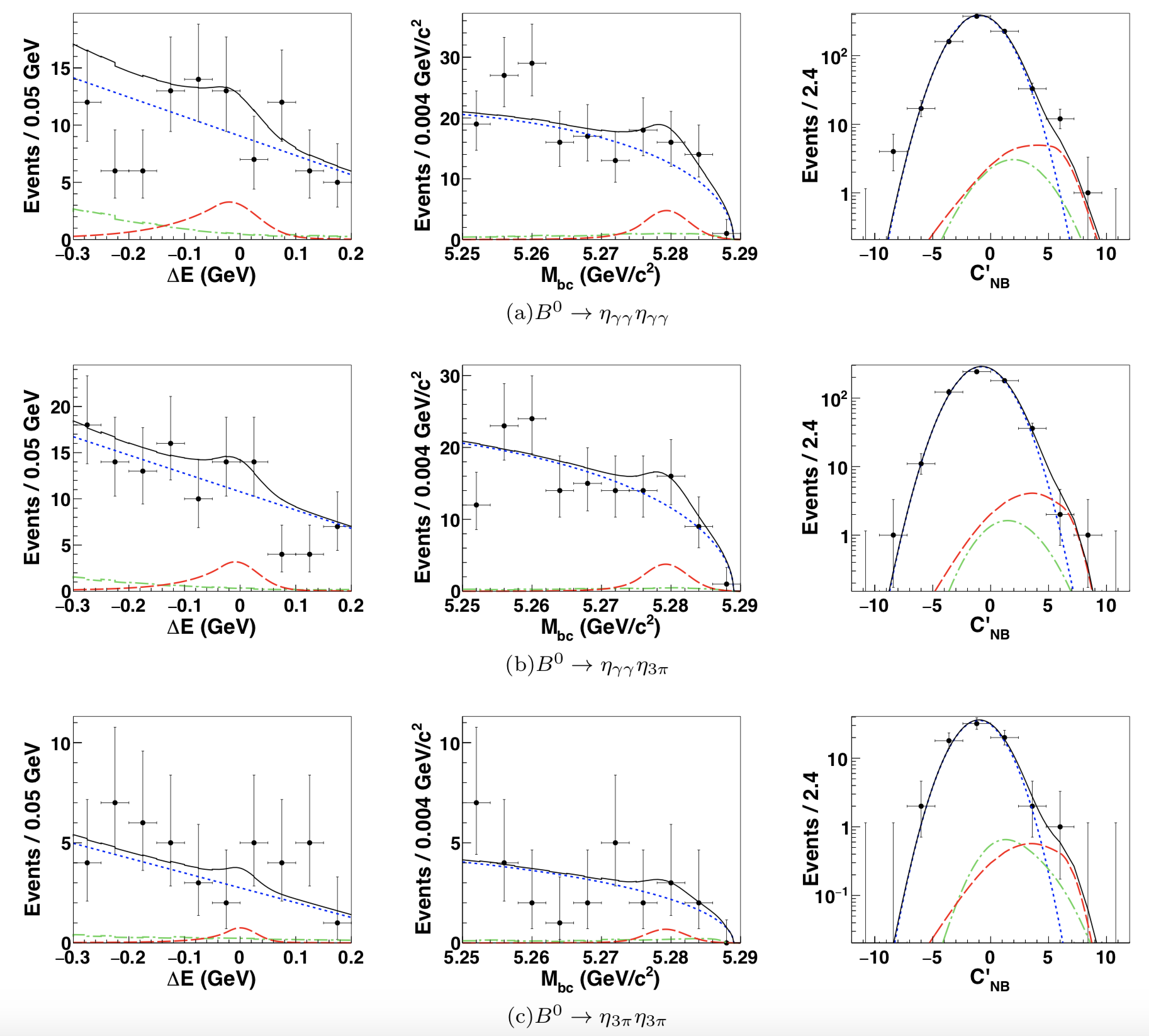}
\caption{\small Signal-enhanced fit projections for the (top) $\Bd\to\eta_{\gamma\gamma}\eta_{\gamma\gamma}$,  (moddle) $\Bd\to\eta_{\gamma\gamma}\eta_{3\pi}$ and (bottom) $\Bd\to\eta_{3\pi}\eta_{3\pi}$ fit results. The points with the error bars show the data; the (black) solid lines show the
total PDF; the (red) dashed lines show the signal; the (blue) dotted lines show the continuum background; and the (green) dot-dashed lines show the other $\BBbar$ background.}
\label{fig:fig3}
\end{figure}

The branching fraction of $\Bd\to \eta \eta$  is obtained by a simultaneous fit to the $\eta_{\gamma\gamma}\eta_{\gamma\gamma}$, $\eta_{\gamma\gamma}\eta_{3\pi}$ and $\eta_{3\pi}\eta_{3\pi}$ decay channels. We perform a three dimensional extended unbinned maximum likelihood fit to the variables $M_{\rm bc}$, $\Delta E$ and  $C'_{\rm NB}$, as shown in Fig.~\ref{fig:fig3}.  We extract $23.6^{+8.1}_{-6.9}$, $9.2^{+3.2}_{-2.7}$ and $2.7^{+0.9}_{-0.8}$ signal events from the $\eta_{\gamma\gamma}\eta_{\gamma\gamma}$, $\eta_{\gamma\gamma}\eta_{3\pi}$ and $\eta_{3\pi}\eta_{3\pi}$ decay channels, respectively.  The measured branching fraction is
\begin{equation*}
\mathcal{B}(\Bd\to\eta\eta) = 
\left( 5.9^{+2.1}_{-1.8}\pm1.4\right) \times 10^{-7},
\end{equation*}
where the first uncertainty is statistical and the second is systematic. The significance of the result
is 3.3 standard deviations, and provides the first evidence of this decay.
\section{\boldmath Branching fraction and \CP asymmetry of $\Bd\to\piz\piz$}
The decay $\Bd\to\piz\piz$ is an important input for the isospin analysis in the $\B\to\pi\pi$ system~\cite{Gronau:1990ka}. 
Among the $\B\to\pi\pi$ decays, this decay is least well determined~\cite{PDG}. 
This decay is also important to probe the disagreement between quantum-chromodynamics-based factorization,
which predicts the branching fraction below $1\times10^{-6}$~\cite{Li:2006cva}, and previous measurements~\cite{PDG} from \belle and \babar.
Here, we present new measurements of $\Bd\to\piz\piz$ based on a $\rm 693 fb^{-1}$ data sample that contains $752\times10^6$ $\BBbar$ pairs~\cite{Julius:2017jso}.

We reconstruct $\Bd\to\piz\piz$ candidates from the subsequent decay of $\piz$ mesons to two photons. In addition to photons reconstructed from electromagnetic calorimeter clusters, which do not match any charged track in the central drift chamber, photons that convert to $\epem$ pairs in the  silicon vertex detector are recovered and reconstructed as $\piz\to\epem\gamma$.  This provides a 5.3\% in crease in detection efficiency.  The dominant background arises from the continuum process. 
To suppress this background, which tend to be
jet-like from spherical $\BBbar$ events, a Fisher discriminant $(T_c)$  is
constructed using the so-called event shape variables. 

The signal yield and  the direct \CP asymmetry ($\mathcal{A}_{CP}$) are extracted via an unbinned extended maximum likelihood fit to the variables 
$M_{\rm bc}$, $\Delta E$ and  $T_c$ in bins of flavor tagging variable.
\begin{figure}[htb]
\centering
\includegraphics[width=0.5\textwidth]{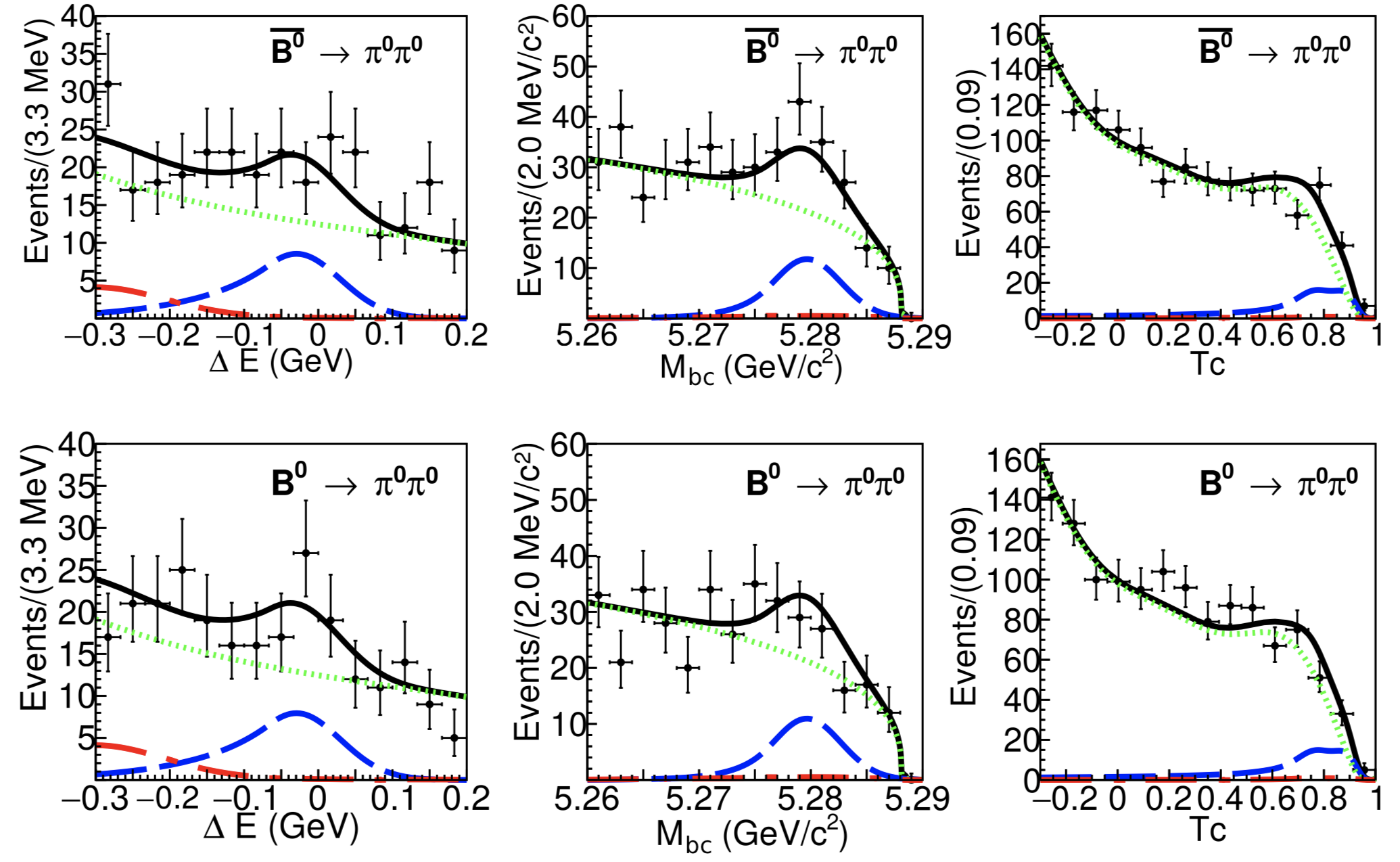}
\caption{\small Projections of the fit results onto (left) $\Delta E$, (middle) $M_{\rm bc}$, (right) $T_c$ are shown in the signal enhanced region: $5.275~\gevcc < M_{\rm bc} < 5.285~\gevcc$, $-0.15~\gev < \Delta E < 0.05~\gev$, and $T_c > 0.7$. Each panel shows the distribution enhanced in the other two variables. Data are points with error bars, and fit results are shown by the solid black curves. Contributions from signal, continuum $\qqbar$, combined $\rho\pi$ and other rare $B$ decays are shown by the dashed (blue), dotted (green), and dash-dotted (red) curves, respectively. The top (bottom) row panels are for events with positive (negative) flavor tagging variable.}
\label{fig:fig4}
\end{figure}

Figure~\ref{fig:fig4} shows the signal-enhanced projections of the fits to data in $M_{\rm bc}$, $\Delta E$ and  $T_c$. We obtain a signal yield of $217\pm32$ events. 
The branching fraction and $\mathcal{A}_{CP}$ are determined to be
\begin{eqnarray*}
\mathcal{B}(\Bd\to\piz\piz)& =&(1.31\pm0.19\pm0.19)\times10^{-6},\\
\mathcal{A}_{CP}&=&+0.14\pm0.36\pm0.10,
\end{eqnarray*}
where the quoted uncertainties are statistical and systematic, respectively. 
Combining our results for $\Bd\to\piz\piz$ with the previous \belle measurements for $\Bd\to\pip\pim$  and $\Bp\to\pip\piz$~\cite{PDG} allows us to employ a isopsin analysis of Ref.~\cite{Gronau:1990ka} to constrain the CKM angle $\phi_2$. Our results exclude $15.5^{\circ} < \phi_2 < 75.0^{\circ}$ at 95\% confidence level.
The measured branching fraction is smaller than our previously published result~\cite{PDG} though consistent within uncertainties. The difference could be due to a substantially smaller fraction of data for which ECL timing information was available (113 of 253 $\rm fb^{-1}$) in the earlier measurement and the subsequent extrapolation to the full data set. The results reported here supersedes our earlier published values and agrees with BaBar measurement~\cite{PDG} within combined uncertainties. While this result is closer to theory predictions than the earlier Belle and BaBar measurements, it is still larger than expectations based on the factorization model~\cite{Zhang:2014bsa}. It is in agreement with the recent works of Qiao {\it et\,al.}~\cite{Qiao:2014lwa} as well as Li and Yu~\cite{Li:2016giw} which employ different theoretical approaches. The upcoming \belle II experiment~\cite{Kou:2018nap}, with its projected factor of 50 increase in luminosity, will enable precision measurements of $\mathcal{B}$ and \CP asymmetry of $\Bd\to\piz\piz$ and other $\B\to\pi\pi$ decays to strongly constrain $\phi_2$.
\section{\boldmath Observation of the decay $\Bs\to K^0\bar{K^0}$}
The two-body decays $B_s^0\rightarrow h^+h'^-$, where $h^{\scriptscriptstyle(}\kern-1pt{}'\kern-1pt{}^{\scriptscriptstyle)}$ is
either a pion or kaon, have now all been observed~\cite{PDG}.
In contrast, the neutral-daughter decays $B_s^0\rightarrow h^0h'^0$ have
yet to be observed. The decay $B_s^0\rightarrow K^0\bar{K^0}$
is of particular interest because the branching fraction is predicted
to be relatively large. In the SM, the decay
proceeds mainly via a $b\rightarrow s$ loop (or ``penguin") transition as shown
in Fig.~\ref{fig:feynman}, and the branching fraction is predicted
to be in the range $(16-27)\times10^{-6}$~\cite{SM-branching}.
The presence of non-SM particles or couplings could enhance 
this value~\cite{Chang:2013hba}. It has been pointed out
that $CP$ asymmetries in $B_s^0\rightarrow K^0\bar{K^0}$ decays are
promising observables in which to search for new
physics~\cite{susy}. 
\begin{figure}[htb]
\centering
\includegraphics[width=0.45\textwidth]{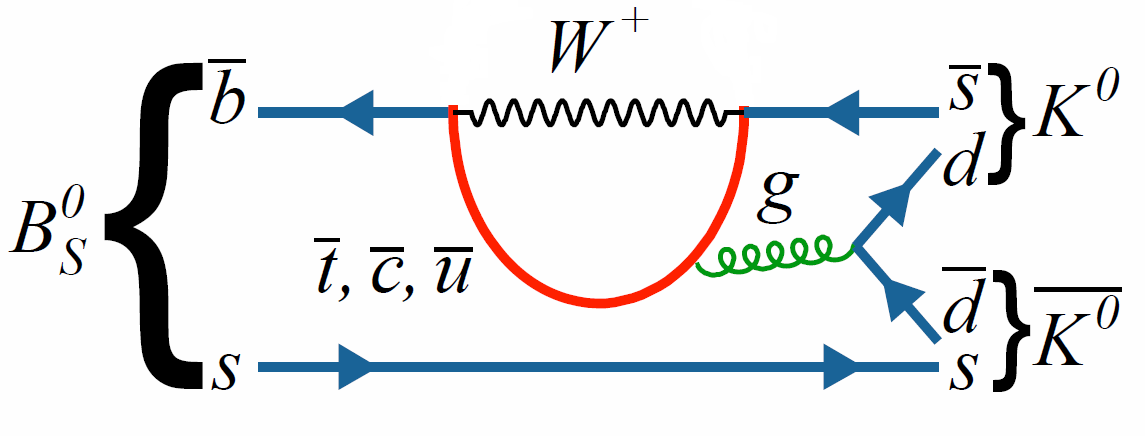}
\caption{\small Loop diagram for $B_s^0\rightarrow K^0\bar{K^0}$ decays.  }
\label{fig:feynman}
\end{figure}

The current upper limit on the branching fraction,
$\mathcal{B}(B_s^0\rightarrow K^0\bar{K^0})<6.6\times 10^{-5}$ at 90\% 
confidence level (C.L.), was set by the Belle Collaboration using 
$23.6~{\rm fb^{-1}}$ of data recorded at the
$\Upsilon(5S)$ resonance~\cite{Peng:2010ze}.
The analysis presented here uses the full data set of
$121.4~{\rm fb^{-1}}$ recorded at the~$\Upsilon(5S)$.
Improved tracking, $K^0$ reconstruction, and continuum suppression algorithms are also used in this analysis. 
The data set corresponds to $(6.53\pm 0.66)\times10^6$  $B_s^0\bar{B_s^0}$
pairs~\cite{Oswald:2015dma} produced in three $\Upsilon(5S)$ decay
channels: $B_s^0\bar{B_s^0}$, $B_s^{*0}\bar{B_s^0}$ or $B_s^0\bar{B}_s^{*0}$, and $B_s^{*0}\bar{B}_s^{*0}$.
The latter two channels dominate, with production fractions
of $f_{B_s^{*0}\bar{B_s^0}}=(7.3\pm1.4)\%$ and $f_{B_s^{*0}\bar{B}_s^{*0}}=(87.0\pm1.7)$\%~\cite{Esen:2012yz}.
The $B_s^{*0}$ decays via $B_s^{*0}\rightarrow B_s^0\gamma$, and the $\gamma$ is not reconstructed.

\begin{figure*}[t]
  \includegraphics[width=0.32\textwidth]{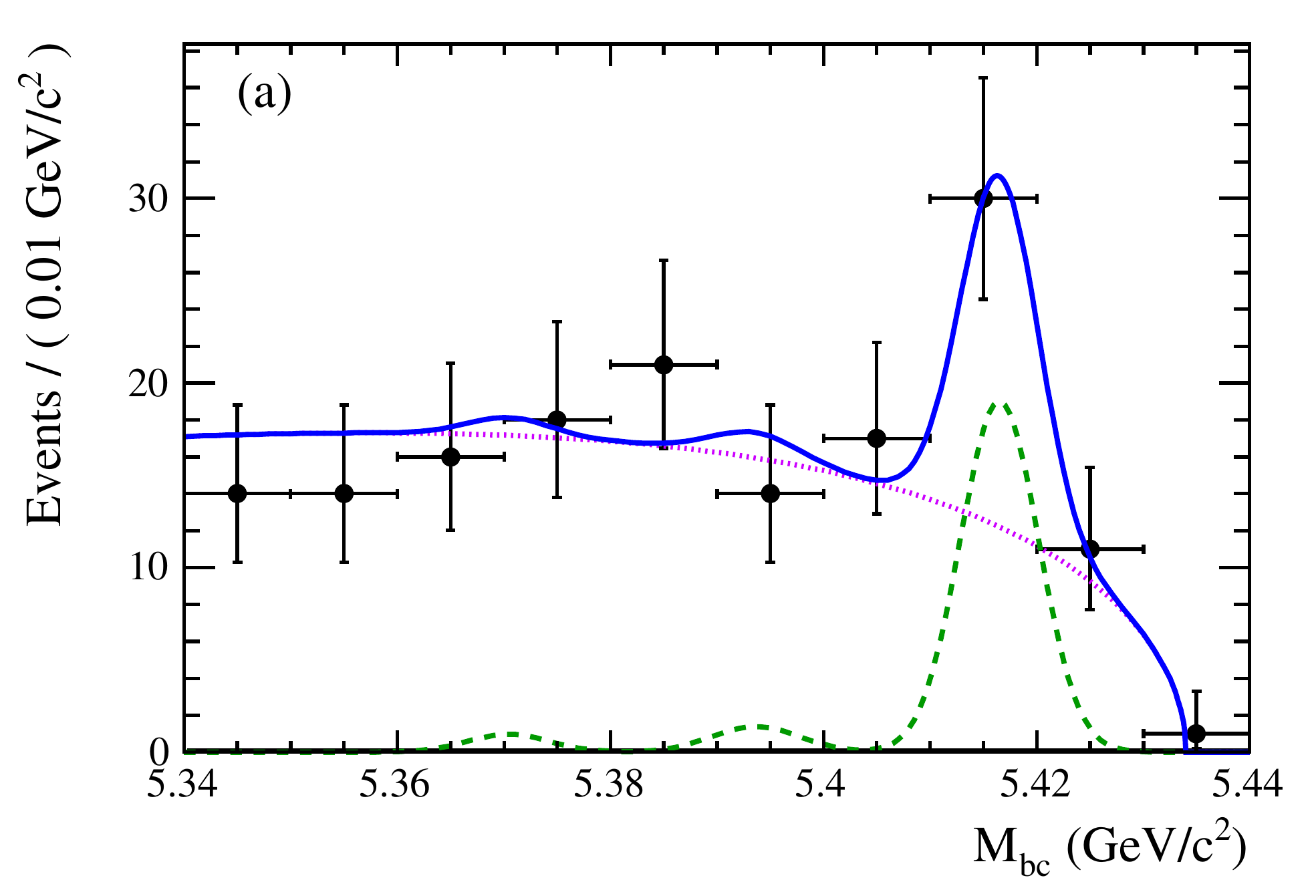}
  \includegraphics[width=0.32\textwidth]{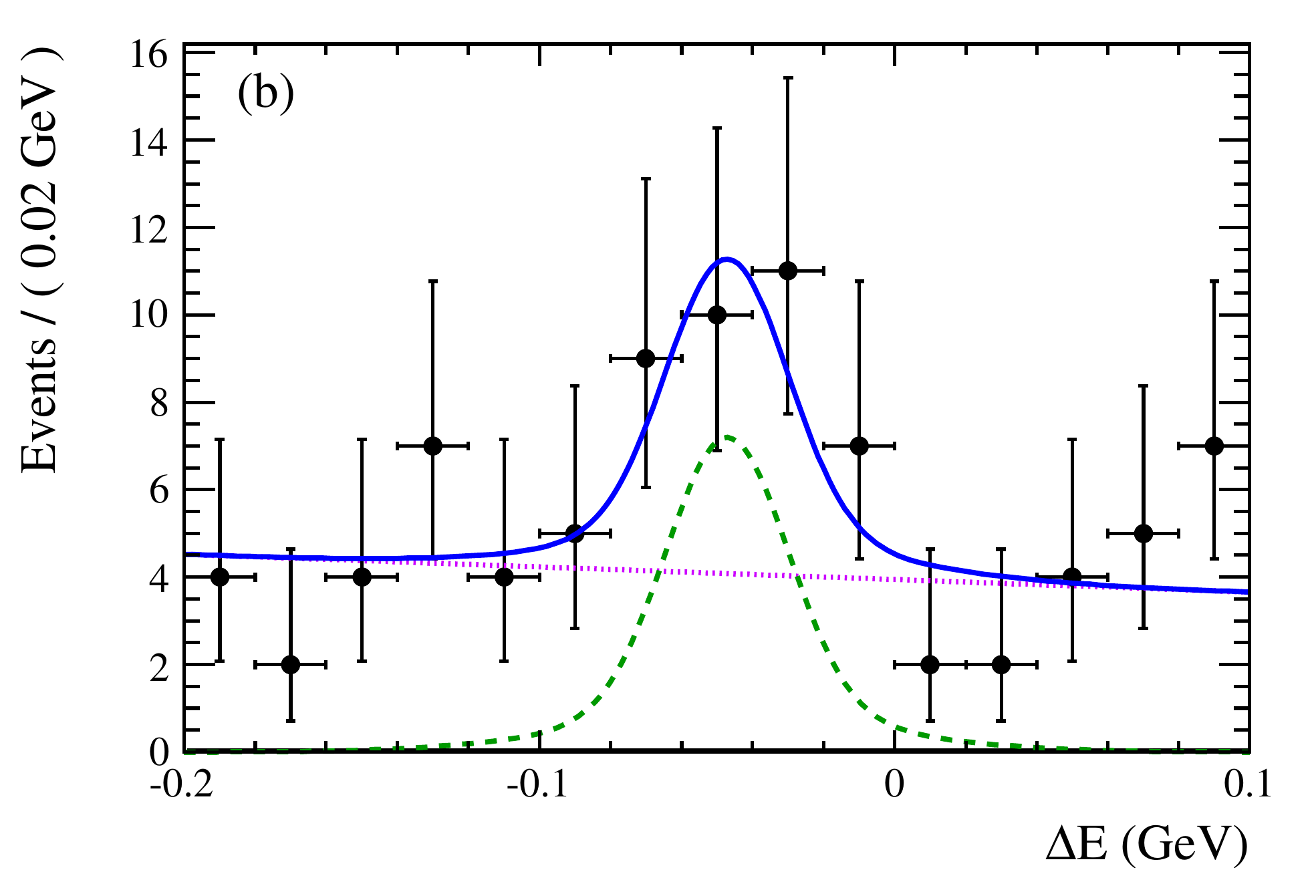}
  \includegraphics[width=0.32\textwidth]{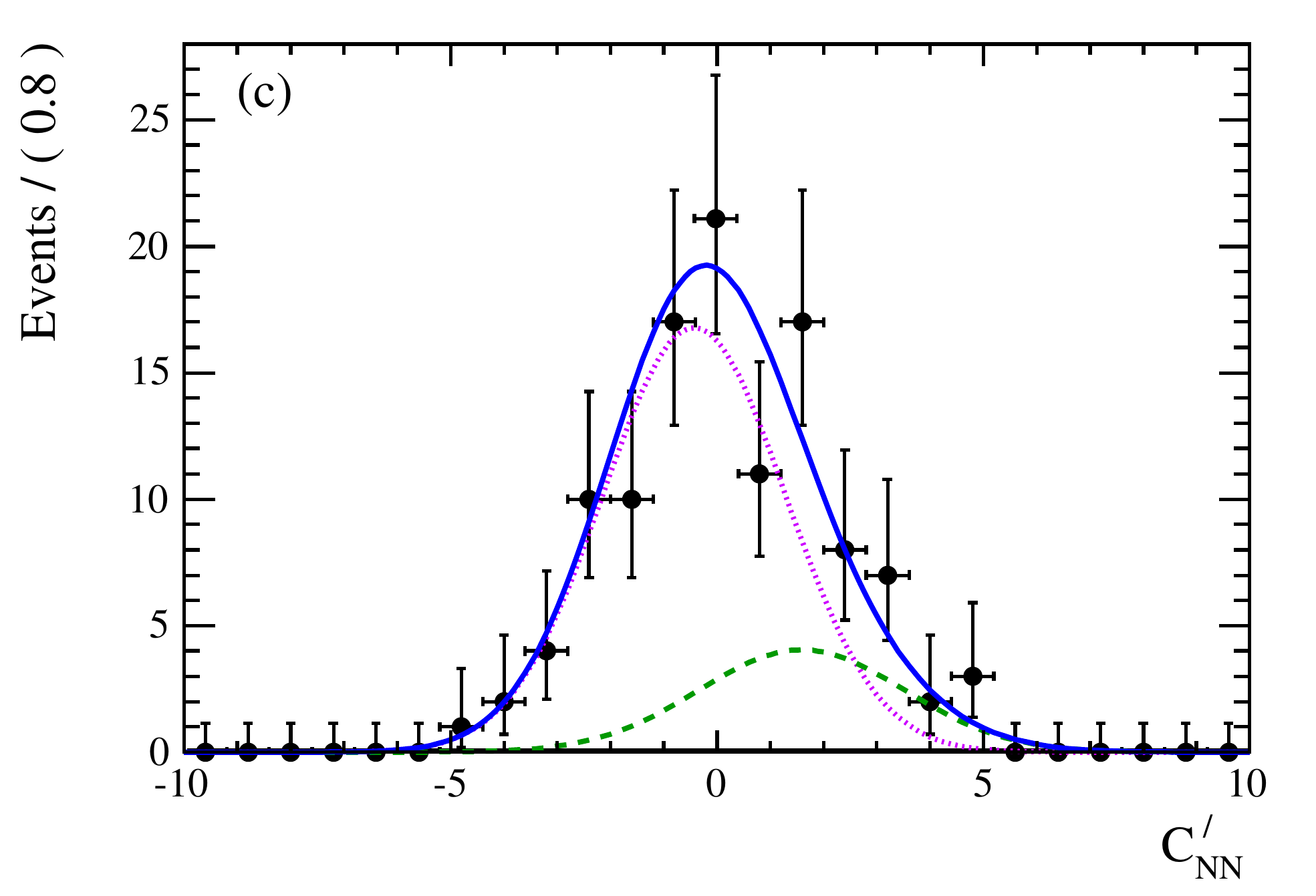}
\caption{\small  Projections of the 3D fit to the real data: 
(a) $M_{\rm bc}$ in $-0.11~{\rm GeV} <\Delta E < 0.02~{\rm GeV}$
and $C^{\prime}_{\rm NN}>0.5$;
(b) $\Delta E$ in $5.405~{\rm GeV}/c^{2} <M_{\rm bc}< 5.427~{\rm GeV}/c^{2}$
and $C^{\prime}_{\rm NN}>0.5$; and 
(c) $C^{\prime}_{\rm NN}$ in $5.405~{\rm GeV}/c^{2} <M_{\rm bc}< 5.427~{\rm GeV}/c^{2}$
and $-0.11~{\rm GeV} <\Delta E < 0.02~{\rm GeV}$. 
The points with  error bars are data, the (green) dashed curves
show the signal, (magenta) dotted curves show the continuum
background, and (blue) solid  curves show the total.  The three peaks in $M_{\rm bc}$ arise from 
$\Upsilon(5S)\to B_s^0\bar{B_s^0}, B_s^{*0}\bar{B_s^0}+B_s^0\bar{B}_s^{*0}$, and $B_s^{*0}\bar{B}_s^{*0}$ decays.
}
\label{fig:fig2}
\end{figure*}

Candidate $K^0$ mesons are reconstructed via the decay $K_S^0\to\pi^+\pi^-$ and require that the $\pi^+\pi^-$ invariant mass be within 12 MeV/$c^2$ of the nominal $K_S^0$ mass~\cite{PDG}. In order to extract the signal yield, we perform a three-dimensional (3D) unbinned maximum likelihood fit to the variables,  $M_{\rm bc}$,
$\Delta E$, and continuum suppression variable $C'_{\rm NN} = \ln\left(\frac{C_{\rm NN}-C^{\rm min}_{\rm NN}}
{C^{\rm max}_{\rm NN}-C_{\rm NN}}\right)$. We extract $29.0\,^{+8.5}_{-7.6}$ signal events
and $1095.0\,^{+33.9}_{-33.4}$ continuum background events.
Projections of the fit are shown in Fig.~\ref{fig:fig2}.
The branching fraction of the decay $B_s^0\rightarrow K^0\bar{K^0}$ is measured to be~\cite{Pal:2015ghq}
\begin{equation}
\mathcal{B}(B_s^0\rightarrow K^0\bar{K^0})=(19.6\,^{+5.8}_{-5.1}\,\pm1.0\,\pm2.0)\times10^{-6},
\end{equation}
where the first uncertainty is statistical, the second
is systematic, and the third reflects the uncertainty due to the 
total number of $B_s^0\bar{B_s^0}$ pairs. The significance of this result is 5.1 standard deviations, thus, our measurement constitutes the first observation of this decay.
This measured branching fraction  is in good agreement with the
SM predictions~\cite{SM-branching}, and it implies that the Belle II experiment~\cite{Kou:2018nap} will
reconstruct over 1000 of these decays. Such a sample would allow for a much higher sensitivity search for new physics in this $b\to s$ penguin-dominated decay.
\section{\boldmath Prospects of  charmless hadronic $B$ decays at \belle II~\cite{Kou:2018nap}}
The large datasets collected by Belle, BaBar, and LHCb have enabled the study of many charmless hadronic $B_{(s)}$ decays and have allowed for a detailed comparison with theoretical predictions and models. Some tantalizing questions have emerged and await the large dataset of Belle II to be further understood. The expected precision in $\Bd\to\KS\piz$ with ${\rm 50~ ab^{-1}}$ of data will be sufficient for NP studies and may resolve the $K\pi$ \CP-puzzle. The analogous isospin sum rules for the multi-body $\pi K^{(*)}$ and $\rho K^{(*)}$  decays are also promising avenues to resolve this puzzle, but are statistically limited and must be measured with high precision to reveal whether an anomalous pattern of direct \CP violation is emerging. The study of $\B\to VV$ decays is still in its infancy due to the large statistics required to perform full angular analyses. While the majority of analyses at \belle and \babar were limited to only measuring the longitudinal polarisation fraction, full angular analyses will be possible for many $VV$ channels at \belle II. Of particular interest are $\rho K^{*}$ decays, where a polarisation analysis will reveal if there is an enhanced contribution proportional to electromagnetic penguins. \belle II will also be uniquely suited to search for \CP asymmetries in $\B\to 3h$ decays with multiple neutral particles in the final state, which will serve to complement related searches at LHCb, where the observation of large local \CP asymmetries in multiple channels has generated enormous interest from the theoretical and phenomenological communities. A size-able $\Bs$ dataset will also be necessary to study rare decays such as the penguin dominated $\Bs\phi\piz$, where an excess above the SM prediction would be a clear indication of NP, and, $e.g.$, the recently observed $\Bs\to K^0\bar{K^0}$ decay, where \belle II expects to reconstruct $\mathcal{O}(1000)$ events with $\rm 5~ab^{-1}$ which will enable a \CP violation study and will serve to clarify the presence of NP in the decay. There are countless additional charmless hadronic $B_{(s)}$ decays which will be within the reach of \belle II. This will open up a new era of discovery and complementarity with other experiments.
\\
\\
\textbf{Acknowledgement}:
The author thanks the workshop organizers for hosting a fruitful and stimulating workshop
and providing excellent hospitality. This research is supported by the U.S. Department of Energy.

\end{document}